\documentclass[iop,apj]{emulateapj}
\usepackage{float}
\usepackage{amsmath}
\usepackage{rotating}
\usepackage{amssymb}
\usepackage{hyperref}

\def\gtrsim{\mathrel{\hbox{\rlap{\hbox{\lower4pt\hbox{$\sim$}}}\hbox{$>$}}}}
\def\lesssim{\mathrel{\hbox{\rlap{\hbox{\lower4pt\hbox{$\sim$}}}\hbox{$<$}}}}
\def\gtrsim{\mathrel{\hbox{\rlap{\hbox{\lower4pt\hbox{$\sim$}}}\hbox{$>$}}}}
\def\farcs{\hbox{$.\!\!^{\prime\prime}$}}
\def\farcm{\hbox{$.\!\!^{\prime}$}}

\begin{document}

\def\chan{{\sl Chandra\ }}

\title{Chandra observations of outflows from PSR J1509--5850}

\author{Noel Klingler$^1$, Oleg Kargaltsev$^1$, Blagoy Rangelov$^1$, George G. Pavlov$^2$, Bettina Posselt$^2$, C.-Y. Ng$^3$}
\affil{$^1$The George Washington University, Department of Physics, 725 21st St NW, Washington, DC  20052 \\ $^2$Pennsylvania State University, Department of Astronomy \& Astrophysics, 525 Davey Laboratory, University Park, PA 16802\\ 
$^3$Department of Physics, the University of Hong Kong, Pokfulam, Hong Kong \\  }

\begin{abstract}
PSR J1509--5850 is a middle-aged pulsar with the period $P\approx 89$ ms, spin-down power $\dot{E}=5.1\times 10^{35}$ erg s$^{-1}$, at a distance of about 3.8 kpc.  We report on deep {\sl Chandra X-ray Observatory} observations of this pulsar and its pulsar wind nebula (PWN). 
In addition to the previously detected tail extending up to $7'$ southwest from the pulsar (the southern outflow), the deep images reveal a similarly long, faint diffuse emission stretched toward the north (the northern outflow) and the fine structure of the compact nebula (CN) in the pulsar vicinity.
The CN is resolved into two lateral tails and one axial tail pointing southwest (a morphology remarkably similar to that of the Geminga PWN), which supports the assumption that the pulsar moves towards the northeast.  The luminosities of the southern and northern outflows are about $1\times 10^{33}$ and $4\times 10^{32}$ erg s$^{-1}$, respectively.
The spectra extracted from four regions of the southern outflow do not show any softening with increasing distance from the pulsar.  The lack of synchrotron cooling suggests a high flow speed or in-situ acceleration of particles.  The spectra extracted from two regions of the northern outflow show a hint of softening with distance from the pulsar, which may indicate slower particle propagation.  
We speculate that the northern outflow is associated with particle leakage from the bow shock apex into the ISM, while the southern outflow represents the tail of the shocked pulsar wind behind the moving pulsar.  We estimate the physical parameters of the observed outflows and compare the J1509--5850 PWN with PWNe of other supersonically moving pulsars.  
\end{abstract}

\keywords{pulsars: individual (PSR J1509--5850) --- stars: neutron --- X-rays: general}

\section{INTRODUCTION}
Rotation-powered pulsars are known to produce relativistic magnetized winds responsible for a substantial fraction of the pulsar's spin-down energy loss.  High angular resolution observations with the {\sl Chandra X-ray Observatory} ({\sl CXO}) have shown that the winds are intrinsically anisotropic (both polar and equatorial outflow components are often seen, with the latter dominating in most cases; see Kargaltsev \& Pavlov 2008). 
The outflow morphology is also influenced by the interaction with the ambient medium, which can introduce further anisotropy.  The motion effects become particularly pronounced when the pulsar speed exceeds the sound speed in the ambient medium leading to the formation of a bow shock. 
In this case, the outflow from the pulsar is expected to acquire a cometary shape with an extended tail behind the pulsar.  Some examples of such outflows are the pulsar wind nebulae (PWNe) associated with PSRs B0355+54 (McGowan et al.\ 2006), J0357--3205 (De Luca et al.\ 2010), and  J1741--2054 (Auchettl et al.\ 2015).
Moreover, there has been growing evidence that PWNe around high-speed pulsars (such as PSR B2224+65, IGR J11014--6103) exhibit puzzling elongated features in addition to cometary tails (Hui \& Becker 2007; Pavan et al.\ 2014, 2015).  To understand the origin of the diverse morphologies of pulsar outflows and their interaction with ambient medium, deep X-ray observations with high spatial resolution are particularly valuable. 

PSR J1509--5850 (J1509 hereafter) is a relatively young pulsar with spin-down age $\tau_{\rm sd}=154$ kyr and energy loss rate $\dot{E}=5.1\times10^{35}$ erg s$^{-1}$ (other parameters are listed in Table \ref{tbl-parameters}).   An X-ray nebula associated with J1509 was discovered in a 40 ks {\sl CXO} observation (Kargaltsev et al.\ 2008; K+08 hereafter). 
It was found to consist of a compact nebula (CN) in the pulsar vicinity and a 6 pc long elongated structure interpreted as a tail behind the pulsar moving in the northeast direction.  The PWN's non-thermal spectra were fit with an absorbed ($N_{\rm H}=2.1^{+0.7}_{-0.5}\times 10^{22}$ cm$^{-2}$) power-law (PL) model with photon indices $\Gamma_{\rm CN}=1.8\pm0.3$ and $\Gamma_{\rm tail}=2.4\pm0.2$ (here and everywhere below the uncertainties are given at 68\% confidence unless specified otherwise).  
The luminosity of the tail was found to be  $1\times10^{33}$ erg s$^{-1} \sim 2\times 10^{-3} \dot{E}$ in the 0.5--8 keV band (at the assumed distance of 3.8 kpc; see Table 1).  Observations with the Australia Telescope Compact Array (ATCA) detected the tail in the radio and revealed a high degree of linear polarization whose orientation suggested a helical magnetic field (Ng et al.\ 2010).  Additionally, a faint H$\alpha$ halo around J1509 and a bow shock cavity were detected (Brownsberger \& Romani 2014).
 
The X-ray spectrum extracted from an $r=0.9''$ aperture around the pulsar position was fitted with an absorbed PL model with a best-fit $\Gamma_{\rm PSR}\approx 2.2$.  Since no X-ray pulsations could be detected (because the 3.2 s frame time was much longer than the pulsar period), the unresolved point-like emission could include a compact PWN contribution.   The pulsed emission was detected in $\gamma$-rays by the {\sl Fermi} Large Area Telescope (LAT) up to energies of 3 GeV, with a PL spectrum ($\Gamma_\gamma = 1.36\pm 0.23$) and high $\gamma$-ray efficiency, $\eta_\gamma \equiv L_\gamma/\dot{E} = 0.34\ (d/3.8 {\rm\ kpc})^2$, higher than those of 75\% of $\gamma$-ray pulsars (see Weltevrede et al.\ 2010 and Abdo et al.\ 2013).

In this paper we report on the results of a series of four deep \chan observations of J1509 and its PWN.  In Section 2 we describe the observations and data reduction.  In Section 3 we present the images and spectral fit results.  In Section 4 we discuss our findings, and in Section 5 we present our conclusions.

\begin{deluxetable}{lc}
\tablecolumns{9}
\tablecaption{Observed and Derived Pulsar Parameters \label{tbl-parameters}}
\tablewidth{0pt}
\tablehead{\colhead{Parameter} & \colhead{Value} }
\startdata
R.A. (J2000.0) & 15 09 27.13  \\
Decl. (J2000.0) & --58 50 56.1  \\
Epoch of position (MJD) & 51463  \\
Galactic longitude (deg) & 319.97  \\
Galactic latitude (deg) & --0.62  \\
Spin period, $P$ (ms) & 88.9  \\
Period derivative, $\dot{P}$ (10$^{-14}$) & 0.92 \\
Dispersion measure, DM (pc cm$^{-3}$) & 137.7  \\
Distance\footnote{\
The distance estimate is based on the dispersion measure and the Galactic electron density distribution the model by Taylor \& Cordes (1993). The model by Cordes \& Lazio (2002) places the pulsar at a distance of 2.6 kpc. However, the smaller distance implies a lower $N_{\rm H}$ than observed, and, based on the J1509's Galactic coordinates, would place the pulsar in between the Sagittarius and Scutum-Centaurus arms, whereas the distance of 3.8 kpc places J1509 in the middle of the Scutum-Centaurus arm. }, 
$d$ (kpc) & 3.8  \\
Distance from the Galactic plane, $z$ (kpc) & 0.04  \\
Surface magnetic field, $B_s$ (10$^{12}$ G) & 0.91  \\
Spin-down power, $\dot{E}$ (10$^{35}$ erg s$^{-1}$) & 5.1  \\
Spin-down age, $\tau_{\rm sd} = P/(2\dot{P})$ (kyr) & 154  \\
\enddata
\tablenotetext{}{Parameters are based on Kramer et al.\ (2003).}
\end{deluxetable}

\section{OBSERVATIONS \& DATA REDUCTION}
We conducted four $\approx$90 ks \chan observations of J1509 over the course of 15 months, from 2013 June 14 (MJD 56457) to 2014 September 23 (MJD 56923), totaling 374 ks (see Table \ref{tbl-obs}).  The target was placed on the ACIS-I detector, operated in Very Faint mode with a 3.2 s time resolution.  We used the \chan Interactive Analysis of Observations (CIAO) software package (ver.\ 4.6) and the {\sl CXO} Calibration Data Base (CALDB, ver.\ 4.6.3) for data processing, and restricted the energy range to 0.5--8 keV for images and spectral analysis.
We created exposure maps for each observation and an exposure-corrected image from the merged observation data using the CIAO routine {\tt merge\_obs}.  CIAO's {\tt wavdetect} tool (a Mexican-hat wavelet source detection routine; Freeman et al.\ 2002) was used to detect field point sources, which were used for image alignment and then subtracted for diffuse emission analysis.
We used the positions of 12 nearby stationary point sources in conjunction with the CIAO tool {\tt wcs\_match}\footnote{http://cxc.harvard.edu/ciao/ahelp/wcs\_match.html} to correct slight discrepancies in the astrometry of the event files.
Spectra were extracted using CIAO's {\tt specextract} tool and analyzed using XSPEC (ver.\ 12.8.2).  In all spectral fits we used the XSPEC {\tt wabs} model with absorption cross sections from Morrison \& McCammon (1983).
To study the morphology of the emission in the immediate vicinity of the pulsar, we simulate the Point Spread Function (PSF) using ChaRT (\chan Ray Trace) ver.\ 2 and MARX ver.\ 5.1.

We also analyzed two {\sl XMM-Newton} observations (ObsIDs 0500630101 and 0500630301) carried out on 2008 January 29 and 2008 March 1.  The EPIC (European Photon Imaging Camera) observations of PSR J1509--5850 employed the pn, MOS1, and MOS2 cameras in full frame imaging mode with the medium filter.   We used the {\sl XMM-Newton} Science Analysis Software (SAS, ver.\ 14.0) for the reprocessing of the observations and data reduction. 
After removal of background flares, the effective exposure times are 45\,ks (pn), 73\,ks (MOS1), and 73\,ks (MOS2). Because of the high background, we only use {\sl XMM-Newton} data to analyze the pulsar spectrum (which, however, is more contaminated by the CN contribution than the {\sl Chandra} spectrum due to the lower spatial resolution of {\sl XMM-Newton}).
To minimize the contamination from the CN and the background, we use circular extraction regions with radii of $10''$ (pn) and $8''$ (MOS1\&2) centered on the pulsar (brightness profile peak).  These extraction regions correspond to  $\approx50$\% of the encircled energy (Figures 6 and 7 in Section 3.2.1.1 of the {\sl XMM-Newton} User Handbook\footnote{http://xmm.esac.esa.int/external/xmm\_user\_support/\\documentation/uhb\_2.1/node17.html}).
We use an $r=45''$ circular background region close to the pulsar (on the same chip) but away from the CN. The resulting net source counts fraction is  $\approx90$\% for all instruments.

\begin{deluxetable}{cccccc}
\tablecolumns{6}
\tablecaption{ Details of \chan ACIS-I Observations \label{tbl-obs} }
\tablewidth{0pt}
\tablehead{\colhead{ObsId} & \colhead{Date} & \colhead{Exposure} & \colhead{Off-Axis} & \colhead{Chip} & \colhead{Roll} \\ \colhead{} & \colhead{} & \colhead{Time [ks]} & \colhead{Angle} & \colhead{No.} & \colhead{Angle} }
\startdata
14523 & 2013  Jun 14  & 93.87 & 2\farcm6 & 2 & 303$^\circ$ \\
14524 & 2013  Sep 15  & 93.87 & 1\farcm7 & 3 & 237$^\circ$ \\
14525 & 2014 May 9 & 92.01 & 1\farcm7 & 3 & 2$^\circ$ \\
14526 & 2014 Sep 23 & 93.86 & 1\farcm7 & 3 & 232$^\circ$ \\
\vspace{-0.3cm}
\enddata
\end{deluxetable}

\section{RESULTS}

\subsection{PWN  Morphology}
The merged image produced from four ACIS-I observations (see Figure \ref{image-raw}) reveals a bright CN around the pulsar (514 net source counts) and two fainter large-scale elongated structures.  In addition to the extended tail (hereafter, southern outflow; 3530 net counts) previously reported by K+08, the much deeper image shows, for the first time, another fainter structure (hereafter, the northern outflow; 1,760 net counts) apparently originating from the pulsar/CN area and visible up to 7$'$ (or 7.7 pc at $d=3.8$ kpc) north of the pulsar.  Below we consider the individual PWN elements in detail.

\subsubsection{Small-scale Structure: Compact Nebula }
\begin{figure*}
\epsscale{1}
\plotone{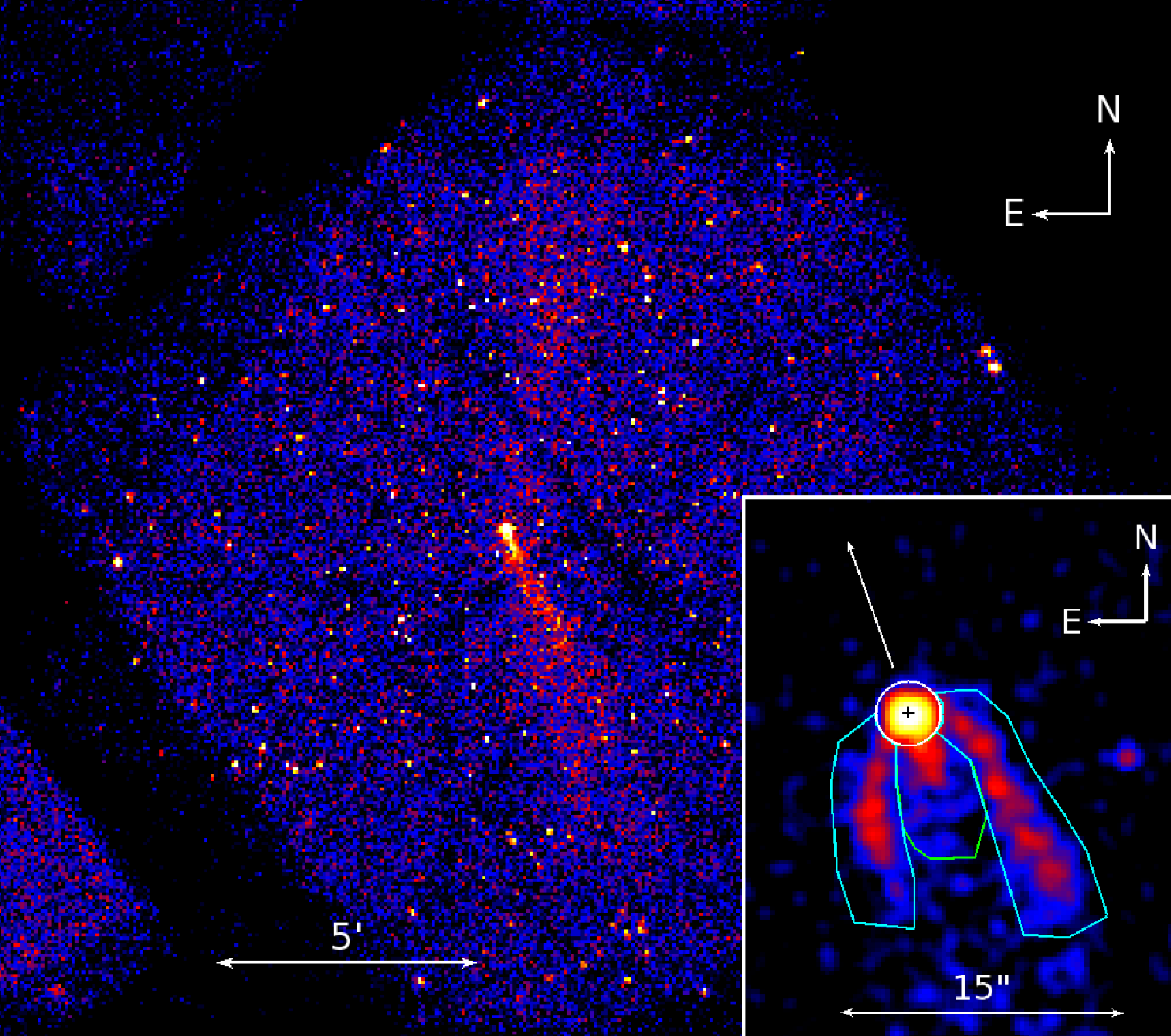}
\caption{Merged image in the 0.5--8 keV band created from the  ACIS-I data from the 4 observations.  The image is binned by a factor of 10 (pixel size $4\farcs9$).  The inset shows the merged image of the CN, binned by a factor of 0.5 (pixel size $0\farcs25$) and smoothed with a $0\farcs74$ Gaussian kernel. The following regions are shown:  pulsar, lateral tails, and axial tail.  The white arrow shows the assumed direction of the pulsar proper motion (inferred from the CN shape; see K+08); the black cross in the $r=1\farcs7$ circle represents the pulsar's radio position.}
\label{image-raw}
\end{figure*}

The inset in Figure \ref{image-raw} shows the CN structure. The peak of the X-ray brightness distribution (R.A.=$15^{\rm h}09^{\rm m}17\fs136$, Decl.=$-58^{\circ}50'56\farcs19$) coincides (within $0\farcs1$) with the radio position of the pulsar (R.A.=$15^{\rm h}09^{\rm m}27\fs13(3)$, Decl.=$-58^{\circ}50'56\farcs1(5)''$ at  MJD 51463, i.e., about 14 years before our X-ray observations).
The centroiding uncertainty of the X-ray position, $\approx0\farcs04$ (calculated by {\tt celldetect}), is much smaller than the typical uncertainty of the {\sl Chandra} absolute astrometry ($\approx0\farcs4$ at 68\% confidence\footnote{http://cxc.harvard.edu/cal/ASPECT/celmon/}). This puts an upper limit of $\sim 0\farcs5$ on the pulsar position shift during 14 years, corresponding to an upper limit $\sim 36$ mas yr$^{-1}$ on proper motion.

The PSF modeling and subtraction does not reveal any additional structure within the core region.  We also attempted PSF deconvolution using the CIAO tool {\tt arestore}, but it did not produce anything new.  The rest of the CN structure bears much resemblance to the Geminga PWN.  Therefore, following Pavlov et al.\ (2006), we will call the three distinct features of the CN the lateral tails and the axial tail\footnote{The naming convention used here does not imply any physical interpretation (see Section 4 for discussion)}.   The lateral tails are noticeably longer ($\approx$7$''-10''$, or 0.13$-$0.18 pc at $d=3.8$ kpc) than the axial tail ($\approx$3$''$, or 0.05 pc at $d=3.8$ kpc).

A wind from a fast-moving pulsar is expected to produce a termination shock with a bow-like (parabolic) structure ahead of the pulsar. However, we see no extended emission ahead of the point-like source at the leading tip of the CN (see inset in Figure 1).  To constrain the stand-off distance $\delta_s$ of the possible unresolved shock apex ahead of the pulsar, we assume that the shape of the lateral tails represent the shock surface, fit the shape with a parabola, and estimate $\delta_s$ as the distance between the vertex of the fitted parabola and the position of the brightness peak.
This approach gave us only upper limits on the stand-off distance, $\delta_s \lesssim 0\farcs3$--1\farcs0, depending on the region chosen for fitting. These fits, however, allowed us to estimate the radius of curvature at the parabola's vertex, $\delta_c\sim 1''$--$2''$.

\begin{figure}
\epsscale{1.15}
\plotone{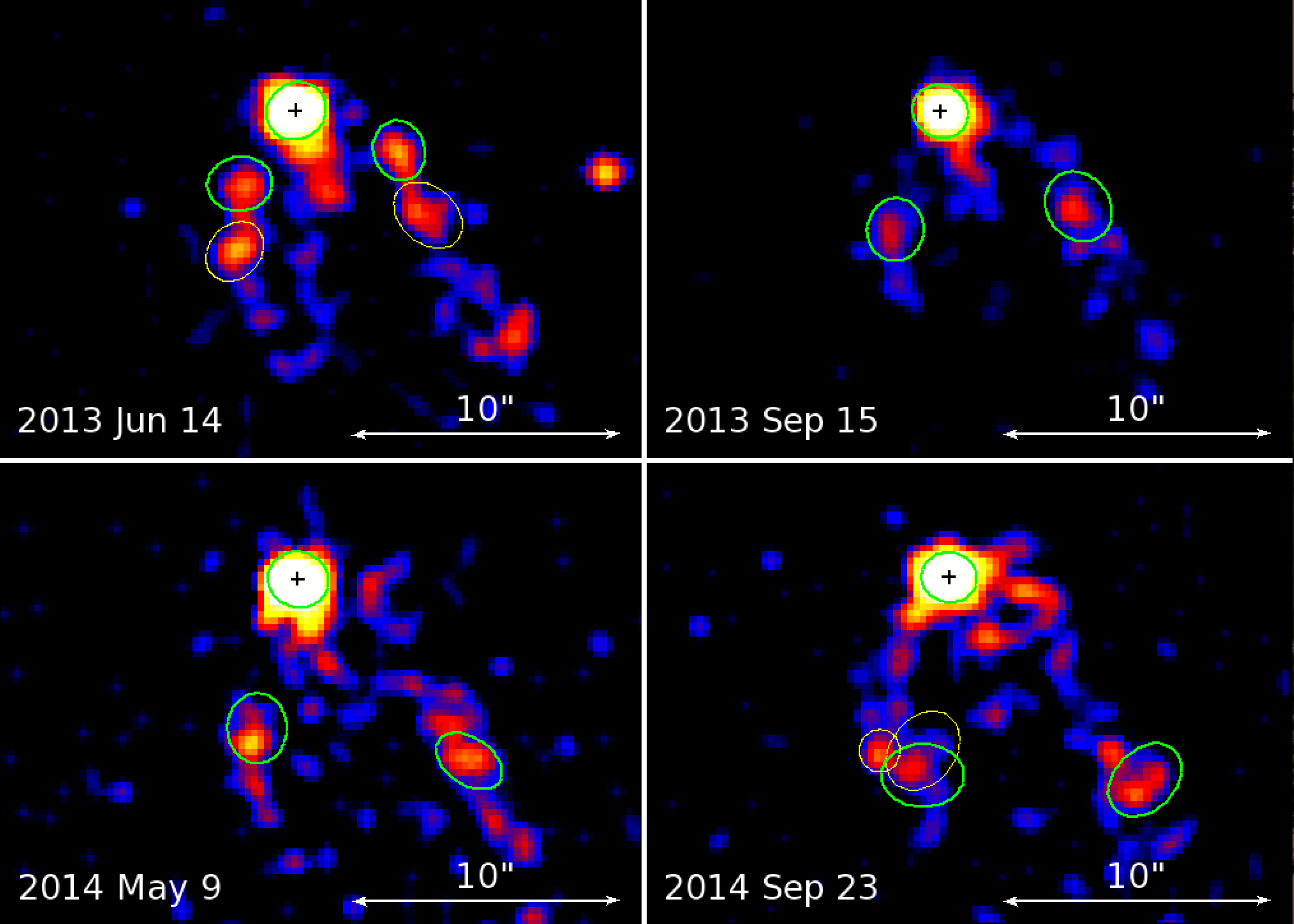}
\caption{Sequence of \chan images showing the apparent changes in J1509's lateral tails. The green ellipses correspond to the blobs participating in the motion fitting (straight lines in Figure \ref{image-blobs-plot}; see also text) while the yellow ones correspond to those also detected by {\tt celldetect} but omitted from the fitting (also shown in Figure \ref{image-blobs-plot}). }
\label{image-blobs-jets}
\end{figure}

\begin{figure}
\epsscale{1}
\plotone{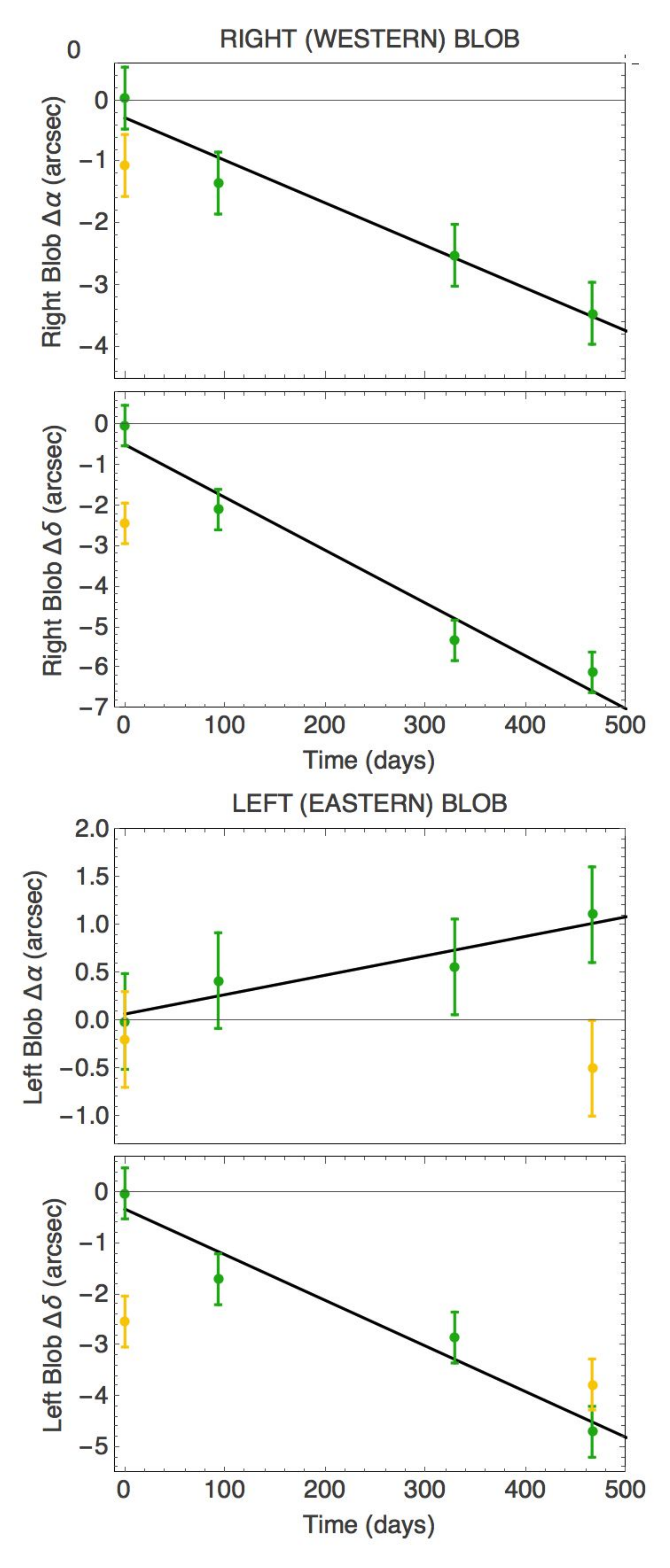}
\caption{Positions of the blobs seen in the western lateral tail (top two plots) and the eastern lateral tail (bottom two plots) for all blobs shown in Figure \ref{image-blobs-jets}.  The lines represent the least-squares fits assuming constant speeds for the green points, which yields an apparent speed of ($0.32\pm0.06$)$d_{3.8}\,c$ for the right blob, and ($0.20\pm0.05$)$d_{3.8}\,c$  for the left blob.  Also plotted in yellow (but not used in the least-squares fits) are the additional regions produced by {\tt celldetect}, shown in Figure \ref{image-blobs-jets}.}
\label{image-blobs-plot}
\end{figure}

Examination of the CN images from individual observations suggests variability associated with the lateral tails.  This variability might be caused by the motion of blobs moving along the tails (see Figure \ref{image-blobs-jets}\footnote{See also the movie at \href{http://home.gwu.edu/~kargaltsev/j1509movie.mov}{http://home.gwu.edu/$\sim$kargaltsev/\\j1509movie.mov})}).
To quantify the changes in the positions of the blobs (shown with green and yellow ellipses in Figure \ref{image-blobs-jets}), we measure their positions (relative to the pulsar) in the images from four separate observations and plot them as a function of time in Figure \ref{image-blobs-plot}. The positions are measured by running CIAO's tool {\tt celldetect}\footnote{The tool also calculates the uncertainties of the blob positions, but they are likely to be underestimated in the presence of fainter emission within the lateral tails (typical $1\sigma$ values range between 0\farcs1--0\farcs2).  Therefore, we conservatively take the actual uncertainties to be 0\farcs5 for all blobs and use these in the fits.}.
It is apparent from these plots that the green data points can be approximately connected by straight lines corresponding to steady motion. If we fit a steady motion model to the green data points (omitting the others from the fitting procedure), the resulting blob speeds are $v_{\rm b,w} = (0.32 \pm 0.06)c$ for the western lateral tail and $v_{\rm b,e} = (0.20 \pm 0.05)c$ for the eastern lateral tail.

This result should be taken with some caution because the significance of the individual blobs in the individual observations is low, $\sim 1.5\sigma - 2\sigma$, when using a local background (i.e. within the lateral outflows) as calculated by {\tt celldetect}).  However, it increases to $3\sigma$--$4\sigma$ when the external (i.e., outside the CN) background region is used.  Moreover, under the assumption of steady motion, the same blob is seen in multiple observations and hence the combined blob significances increase to 3$\sigma$ for the local background choice, and to $7\sigma$--$8\sigma$ for the external background choice. 

We have also compared the merged image of the CN with the older observation (ObsID 3515, ACIS-S, 2003 Feb 9).  That observation was significantly shorter (40 ks), and the ACIS-S3 background contribution was much higher than that of the ACIS-I, making it difficult to  resolve the individual components of the CN (see Figure 4 in K+08).

\begin{figure}
\epsscale{1.15}
\plotone{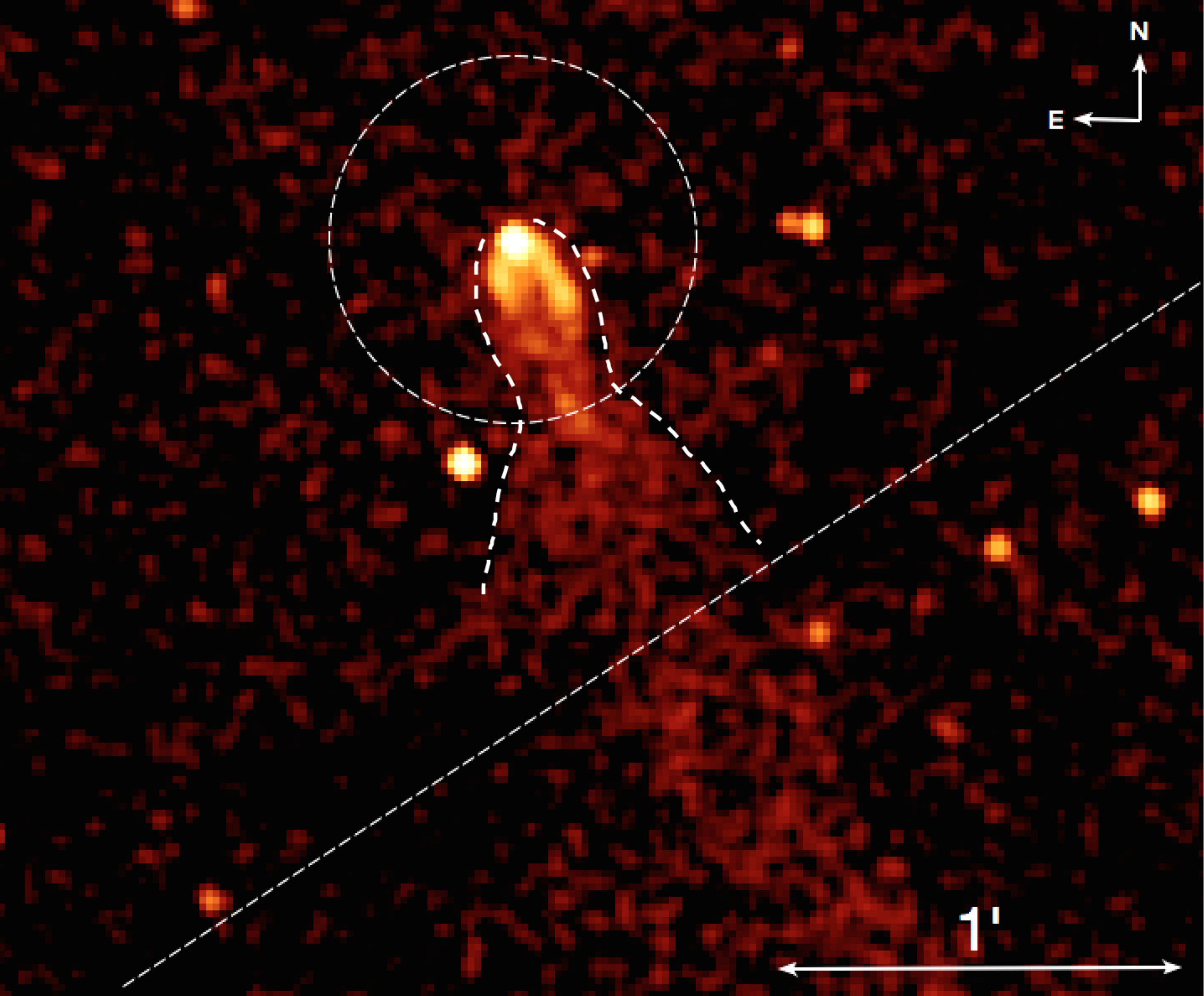}
\caption{ACIS-I image demonstrating the transition from the CN to the southern outflow.  The image is binned to $1''$ pixels and smoothed with an $r=2''$ Gaussian kernel.  The dashed circle of $28''$ radius  shows the size of the H$\alpha$ halo shown in Figure 5 of Brownsberger \& Romani (2014). The curved lines outline the shape of the faint X-ray emission seen in this image (they also correlate with an H$\alpha$ bow shock cavity).  The straight dashed line indicates the ACIS-I array chip gap in one of the observations.}
\label{image-mid-scale}
\end{figure}

\subsubsection{Large-scale Structure}
 
\begin{figure}
\epsscale{1.1}
\plotone{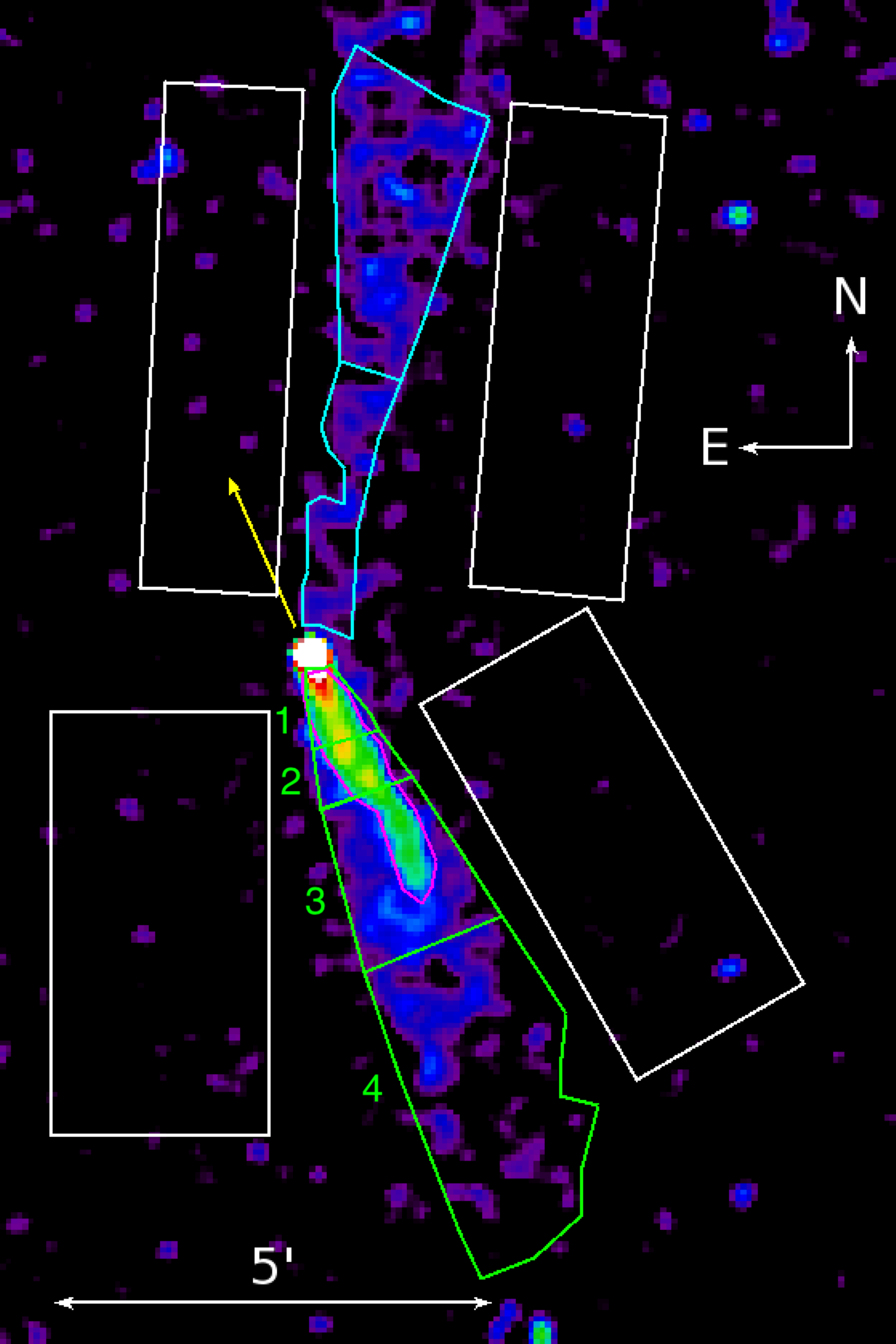}
\caption{Exposure-map-corrected, merged {\sl Chandra} ACIS image showing the large-scale structure of the J1509 PWN in the 0.5--8.0 keV band. The image is binned by a factor of 8 (pixel size $3\farcs9$) and smoothed with an $11\farcs8$ Gaussian kernel, with most point sources (those detected by {\tt wavdetect} removed).  The following regions are shown: northern outflow (green northern bisected polygon), southern outflow (entire large green southern polygon), southern outflow segments 1--4 (numbered green polygon segments), and the spine (magenta polygon, inside of the southern outflow).  The direction of the yellow arrow shows the presumed direction of the pulsar's proper motion (i.e., the CN symmetry axis).  The background regions used for spectral extraction are shown with white boxes.}
\label{image-large-scale}
\end{figure}

Figure \ref{image-mid-scale} shows the transition from the CN to the extended outflow south of the pulsar.  One can see the lateral tails fading and apparently merging at $\approx$27$''$ from the pulsar.  Beyond this distance the flow expands in the transverse direction and gives rise to the fainter {\em southern outflow} seen in Figure \ref{image-large-scale} that presents the merged image of the large-scale diffuse emission with field point sources subtracted and exposure-map correction applied\footnote{To create this image, we followed the procedure described at http://cxc.harvard.edu/ciao/threads/merge\_all/}. 
Unexpectedly, the same image clearly reveals a similarly shaped emission north-northwest of the pulsar, although this {\em northern outflow} is not discernible in the pulsar vicinity (Figure \ref{image-mid-scale}).

\begin{figure}
\epsscale{0.9}
\plotone{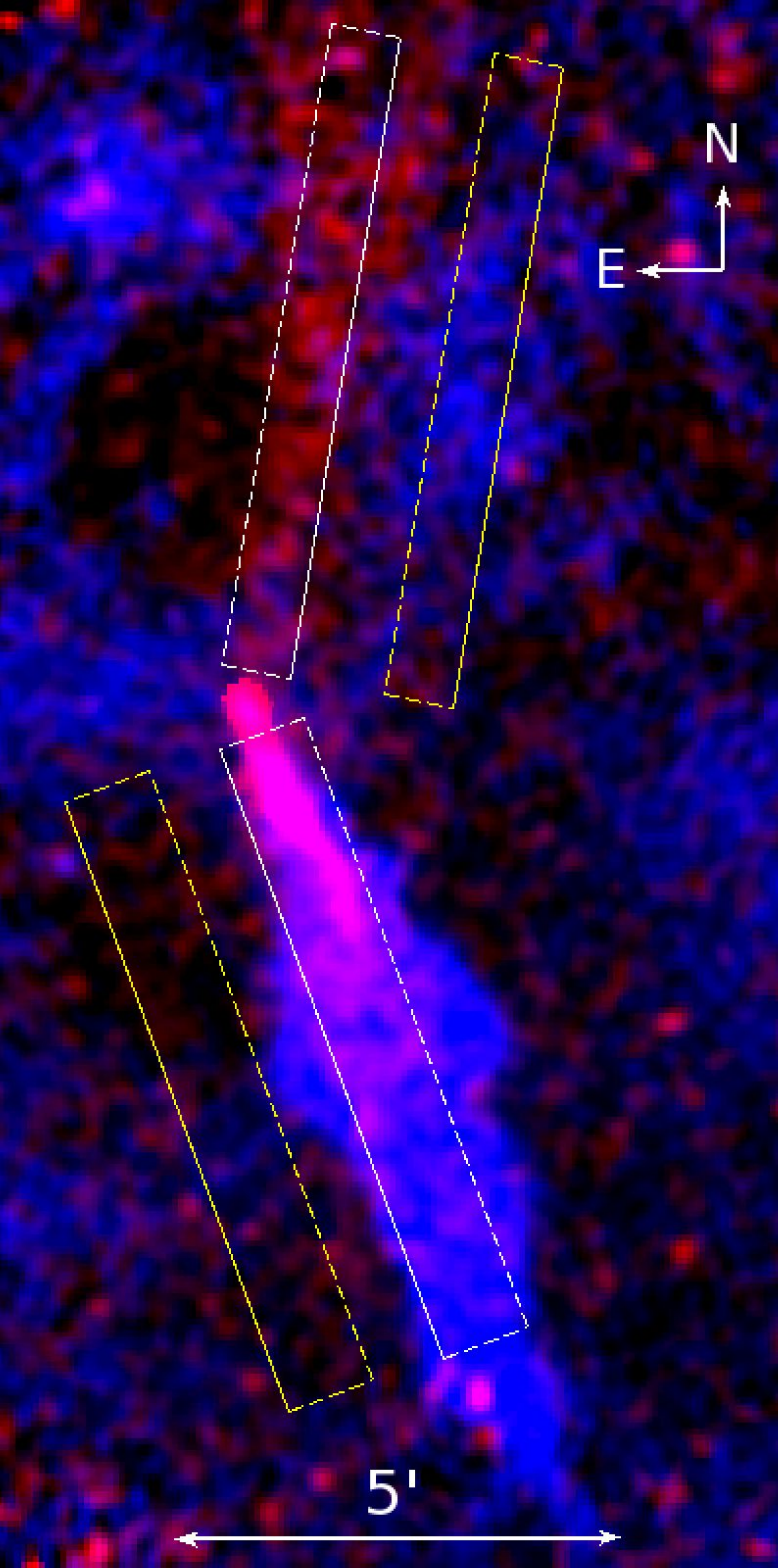}
\caption{Composite image produced from the merged, exposure-map-corrected (fluxed) \chan ACIS images (red; binned by a factor of 8) and the 6 cm  ATCA image (blue; see Ng et al.\ 2010) showing the extended emissions of J1509, both smoothed with a $r=11\farcs8$ Gaussian kernel.  The white boxes show the regions used to obtain the X-ray brightness profiles (southern $7\farcs3\times 1'$ and northern $7\farcs3\times0\farcs8$ outflows), and the yellow boxes (to the sides of the outflows) represent the regions used to measure the background (see Figures \ref{figure-tail-brightness} and \ref{figure-headlight-brightness}).}
\label{image-radio}
\end{figure}

\begin{figure}
\epsscale{1.15}
\plotone{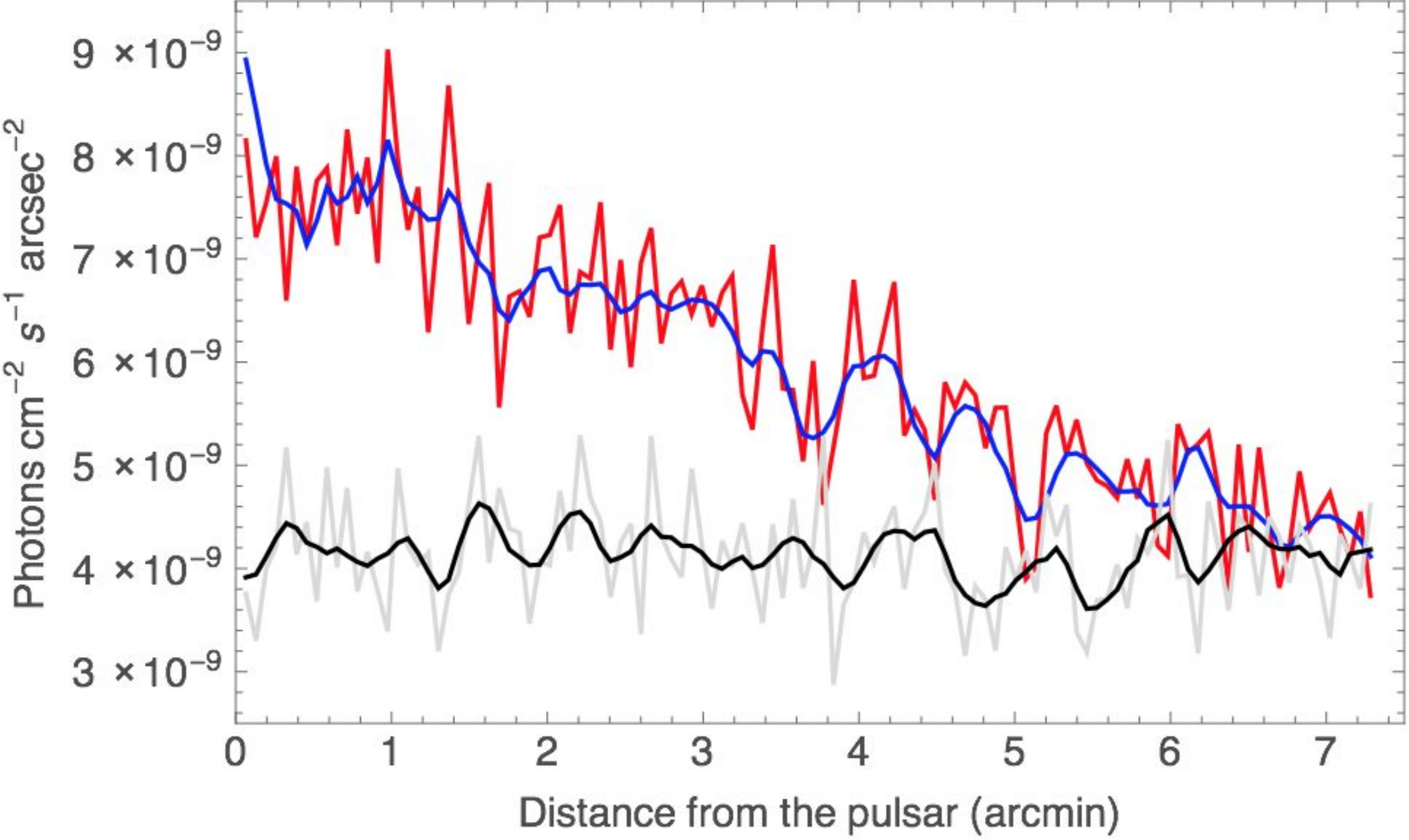}
\caption{Brightness profiles along the southern outflow.  The profiles were obtained from the 0.5--8 keV exposure-map-corrected ACIS-I image binned by a factor of 8 (corresponding to a pixel size of $3\farcs9$). The red and gray curves show the unsmoothed profiles in the southern white and yellow (background) boxes depicted in Figure \ref{image-radio}, while the blue and black curves show the corresponding profiles additionally smoothed with a $11''$ Gaussian kernel.}
\label{figure-tail-brightness}
\end{figure}

\begin{figure}
\epsscale{1.15}
\plotone{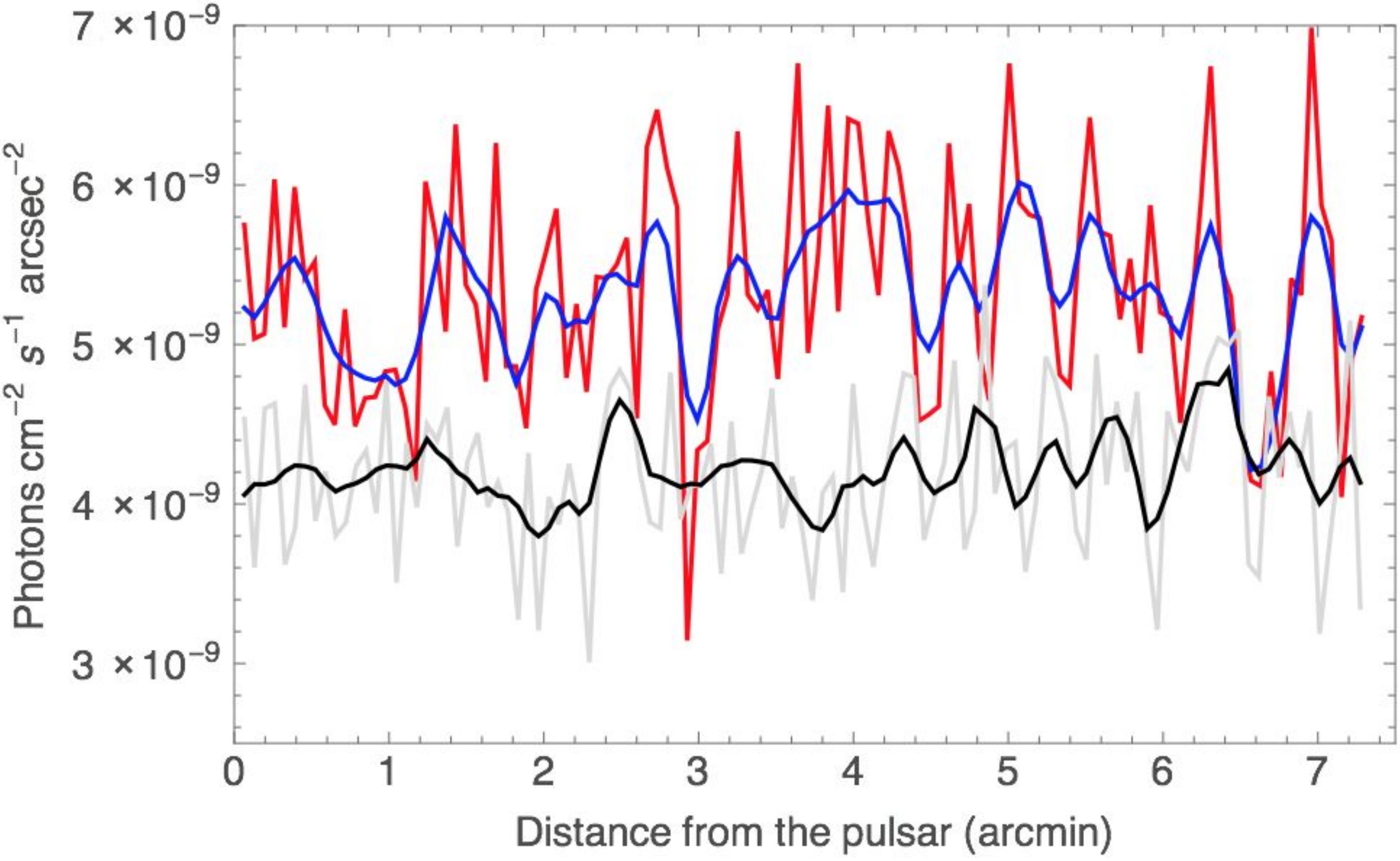}
\caption{X-ray brightness profiles of the northern outflow obtained similarly to the profiles shown in Figure \ref{figure-tail-brightness}.}
\label{figure-headlight-brightness}
\end{figure}

The shapes of both large-scale structures, extending up to $6'-7'$ from the pulsar, can be crudely described as two wedges with $\approx 17^\circ$ opening angles and tips at (or near) the pulsar position.  The southern outflow is approximately aligned with the CN symmetry axis, while the northern outflow is inclined to the CN symmetry axis by $\approx 33^\circ$.
Although the sizes and shapes of these outflows are similar, the distributions of surface brightness within the two structures are markedly different.  The averaged (over the transverse direction) surface brightness of the southern outflow (see Figure \ref{image-radio} for the profile extraction regions) appears to decrease almost linearly with distance from the pulsar (see Figure \ref{figure-tail-brightness}), while the averaged surface brightness of the northern outflow does not appear to change significantly over its length (shown in Figure \ref{figure-headlight-brightness}).  
Moreover, a substantially brighter region (hereafter referred to as the ``spine'') is clearly seen within the southern outflow up to $\approx 4'$ from the pulsar, while no such a feature is seen in the northern outflow. 

Figure \ref{image-radio} also shows the extended radio emission reported by Ng et al.~(2010), which stretches along the southern outflow.  It becomes wider than the X-ray outflow beyond $\approx 3'$ from the pulsar and then narrows down again beyond $\approx 5'$.  The radio outflow is seen up to $10'$ from of the pulsar, whereas the visible southern X-ray outflow becomes comparable with the background at about $7'$ from the pulsar.   No radio emission is seen from the northern outflow region in the same ATCA observation.
 
\subsection{Spatially-resolved Spectra}
We measure the CN spectra in three regions shown in the Figure \ref{image-raw} inset: (1) the bright pulsar/head region (within the $1\farcs7$ radius white circle around the pulsar), which contains an unresolved CN ``head'' in addition to the pulsar, (2) the lateral tails (cyan polygons), and (3) the axial tail (the green polygon).
For the southern outflow, we define the following regions (shown in Figure \ref{image-large-scale}): the entire southern outflow (combination of green polygons), the spine (magenta polygon), and the southern outflow excluding the spine.  We also divide the entire southern outflow region into 4 segments, with segment 1 being the closest to the pulsar and segment 4 being the farthest.  
For the northern outflow, we define two regions: the proximal and distal halves (cyan polygons).  We fit all spectra with an absorbed PL model (unless specified otherwise) and list the fitting parameters Table \ref{spatially-resolved-spectra}.  For all regions, the spectra from different observations are fitted simultaneously, with the all parameters being tied together.
Fitting the spine of the southern outflow yields an absorbing column density $N_{\rm H}= (1.66\pm0.19)\times 10^{22} $  cm$^{-2}$ and a photon index $\Gamma=2.11\pm0.07$, with reduced chi-squared $\chi_\nu^2=1.01$ for $\nu=73$ degrees of freedom (d.o.f.).  Since the spectrum from this region fits the PL model well, and has a relatively small background contribution and a good statistic, we fix $N_{\rm H}$ at the best-fit value obtained for this region while fitting the spectra from other regions which have poorer statistics\footnote{We have also obtained the same value of $N_H$ by fitting the spectra from the lateral tails and the pulsar/head region.}. 
 
For the CN regions, we obtained very similar PL slopes for the pulsar/head and the lateral tails ($\Gamma=1.90\pm0.12$ and $1.80\pm0.13$, respectively) while the axial tail appears to have a harder spectrum ($\Gamma=1.43\pm0.18$).  It is plausible that a substantial fraction of the pulsar/head emission comes from the neutron star and hence it may have a thermal component.
Therefore, we attempted to fit the pulsar/head spectrum with a PL + blackbody (BB) model (with a fixed BB radius of 12 km).  This resulted in a slightly better fit, with $\Gamma=1.77\pm0.13$, $kT=103\pm4$ eV, and $\chi_\nu^2=0.77$ for 53 d.o.f.  However, the best-fit BB+PL model is preferred over the PL model at a 93\% ($1.8\sigma$) confidence, according to the F-test (i.e., the BB component is not required by the data).  
We find an upper limit on the neutron star surface temperature by gradually increasing the temperature of the BB component (and re-fitting the PL component at each new temperature chosen) until the model becomes inconsistent with the data at a $3\sigma$ confidence level.  This yields an upper limit $kT=115$ eV (or $T=1.3\times10^6$ K). 

\begin{figure}
\epsscale{1.2}
\plotone{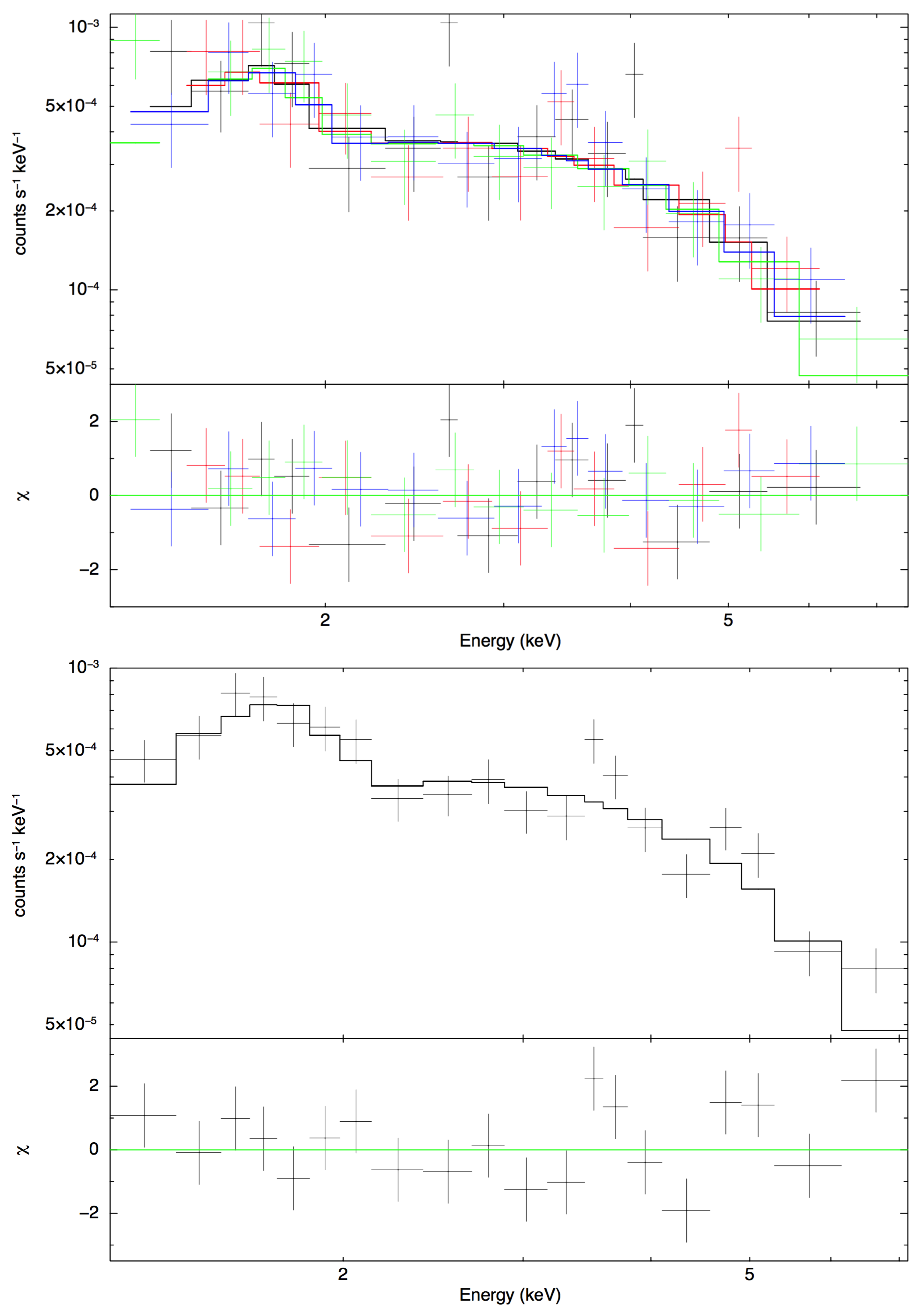}
\caption{The ACIS-I spectrum for the pulsar/head region.  {\sl Top:} Simultaneous fitting of the spectra from the 4 observations with an absorbed PL model ($\Gamma= 1.90\pm0.12$, $N_{\rm H}=1.66$, and $\chi_\nu^2=0.82$ for 54 d.o.f.).  {\sl Bottom:} Pulsar spectrum obtained by merging the spectra from the 4 observations. The best-fit shown has parameters  similar to those obtained from simultaneous fitting.}
\label{image-psr-spectra}
\end{figure}

\begin{figure}
\epsscale{1.2}
\plotone{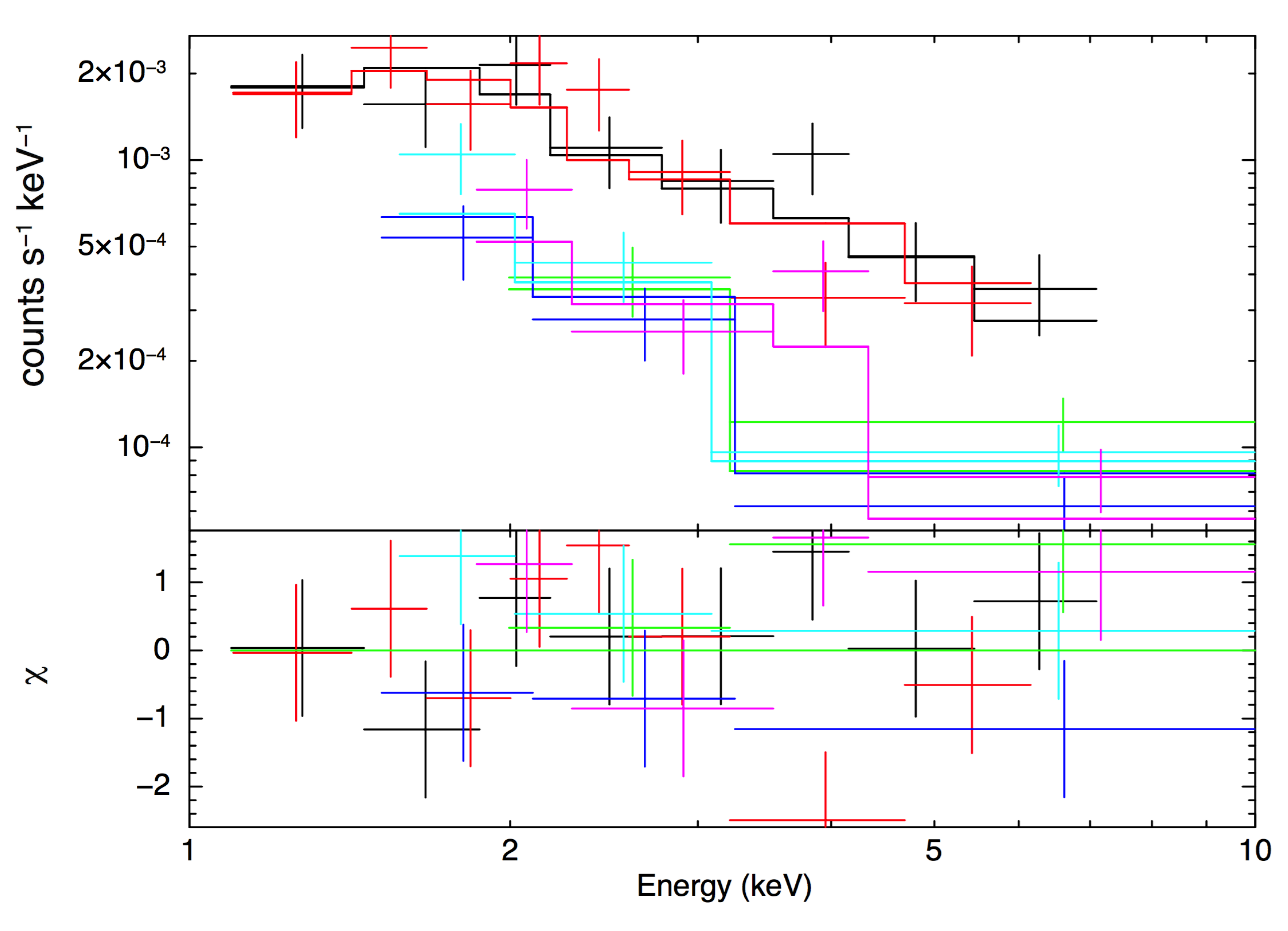}
\caption{Simultaneous fits of the pulsar spectra obtained from both archival {\sl XMM-Newton} observations.  Different observations and instruments are coded with different colors: black --  1st observation pn,  red -- 2nd observation pn, green -- 1st observation MOS1, blue -- 2nd observation MOS1, light blue - 1st observation MOS2, magenta  -- 2nd observation MOS2. Also shown is the best-fit PL model with fixed $N_{\rm H} = 1.66\times 10^{22}$ cm$^{-2}$, $\Gamma=1.9\pm0.2$, and observed flux of $(1.1\pm0.1)\times 10^{-13}$ erg cm$^{-2}$ s$^{-1}$ (in 0.5--10 keV; not corrected for the finite extraction aperture size).}
\label{image-xmm-psr-spectra}
\end{figure}

Although the quality of the fits is good for all the CN regions (see Table \ref{spatially-resolved-spectra}), the fits of the pulsar/head spectrum (see Figure \ref{image-psr-spectra}) shows some systematic residuals centered at 3.5 keV (between 3.3--3.6 keV).  To examine plausibility of the alleged spectral feature, we jointly fit the {\sl XMM-Newton} spectra from two observations (see Figure \ref{image-xmm-psr-spectra}) extracted from  circular regions ($r=10''$ and $8''$ for the pn and MOS detectors; see Section 2) centered on the pulsar with the absorbed PL model and $N_{\rm H}$  fixed at the same value as for the ACIS spectral fits. 
This gave a similar spectral slope, $\Gamma=1.9 \pm 0.2$, but a hint of an emission feature at $\approx 3.5$ keV is seen in the pn spectrum from only one observation while the pn spectrum from the other observation shows a hint of absorption, which means that the feature is likely a statistical fluctuation. The MOS spectra do not have enough counts to look for spectral features. 

\begin{figure}
\epsscale{1.2}
\plotone{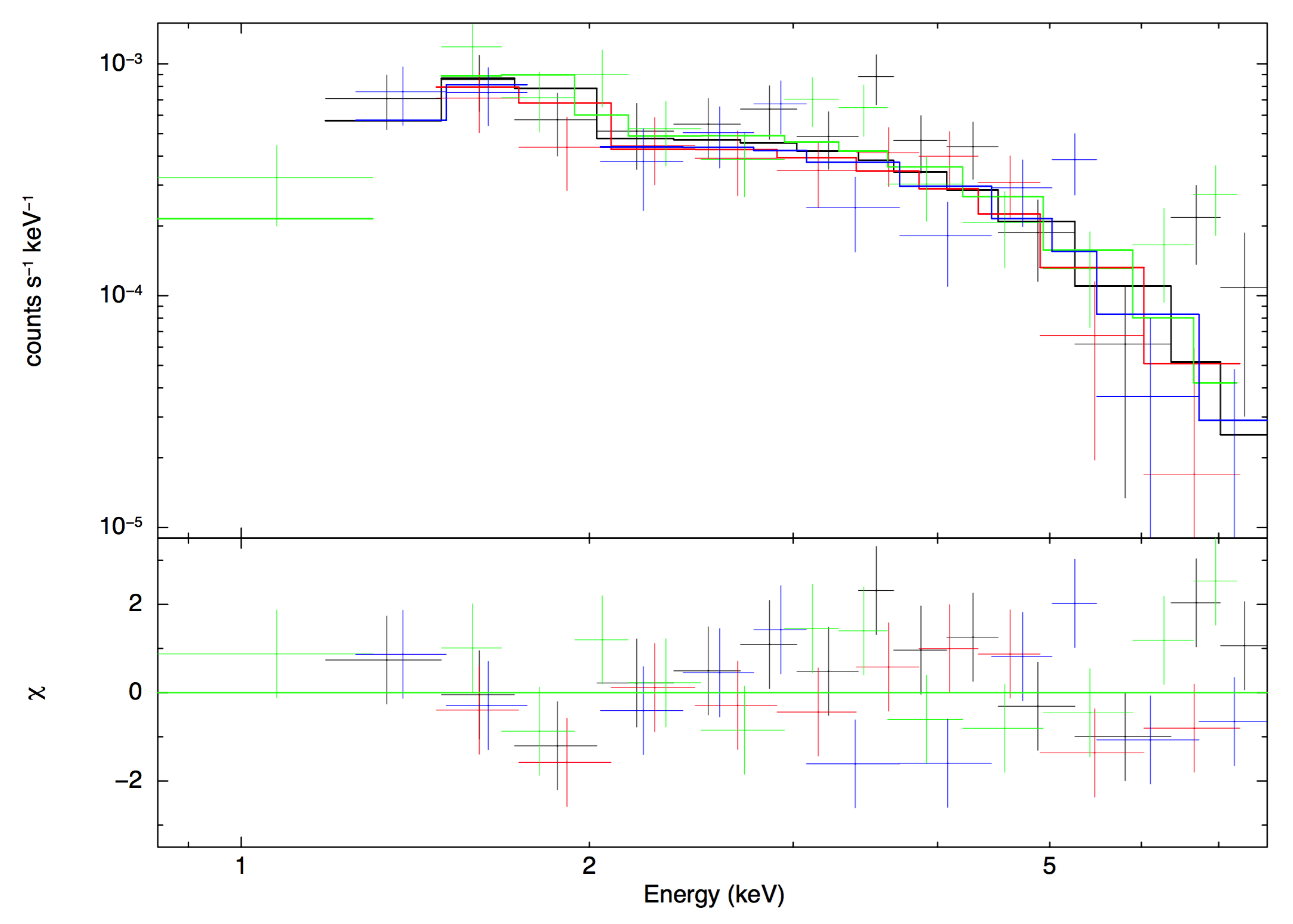}
\caption{The ACIS-I spectrum for the southern outflow segment 1 obtained by simultaneously fitting the spectra from the 4 observations with an absorbed PL model (see Table \ref{spatially-resolved-spectra} for the fitting parameters).}
\label{image-tail-seg-1-spectra}
\end{figure}

The spectrum of the spine ($\Gamma=2.11\pm0.07$) appears somewhat softer than that of the rest of the southern outflow  ($\Gamma=1.78\pm0.09$).  For the southern outflow segments 1--4, we obtain $\Gamma_{\rm S1}=1.99\pm0.12$, $\Gamma_{\rm S2}=2.28\pm0.16$, $\Gamma_{\rm S3}=2.06\pm0.10$, and $\Gamma_{\rm S4}=1.82\pm0.12$.  We do not see any systematic softening of the spectrum along the southern emission despite its large extent; rather there is a hint of hardening from segment 2 to 4.   Almost all fits are of good quality (the poorest quality fit, $\chi_\nu^2=1.26$ for 46 d.o.f., being for segment 1, its spectrum is shown in Figure \ref{image-tail-seg-1-spectra}). 

\begin{figure}
\epsscale{1.2}
\plotone{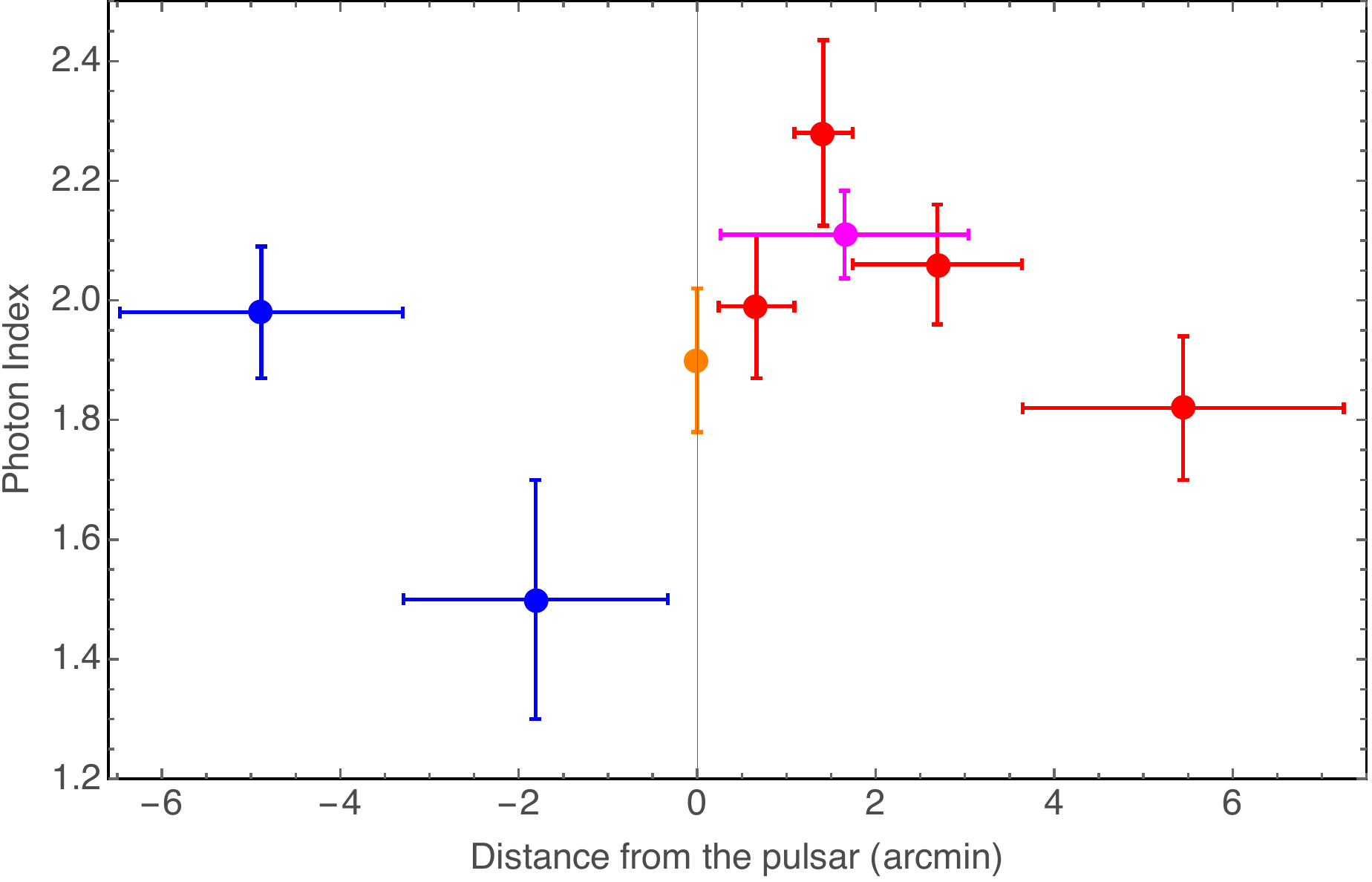}
\caption{Photon index as a function of distance from the pulsar.  From the left, the points represent the following regions: the distal and proximal halves of the northern outflow (blue), pulsar/head (orange), southern outflow segments 1 and 2 (red), spine (magenta), and southern outflow segments 3 and 4 (red).  The horizontal bars indicate the distance over which the photon index was measured.  The vertical bars are $1\sigma$ error bars.}
\label{image-gamma-fn-dist}
\end{figure}

The slope of the spectrum for the entire northern outflow ($\Gamma=1.81\pm0.10$) is similar to the spectrum of the southern outflow excluding the spine ($\Gamma=1.78\pm0.09$) although, unlike the southern outflow, the northern outflow spectrum shows a hint of softening with increasing distance from the pulsar ($\Gamma$ changes from $1.50\pm0.20$ to $1.98 \pm0.11$). This may explain the somewhat poorer quality of the PL fit to the spectrum of the entire northern outflow (see Table \ref{spatially-resolved-spectra}).  The photon indices for all regions as a function of distance from the pulsar are shown in Figure \ref{image-gamma-fn-dist}.

\begin{deluxetable*}{lrccccccc}[H]
\tablecolumns{4}
\tablecaption{Spectral Fits for Different Regions
\label{spatially-resolved-spectra}}
\tablewidth{0pt}
\tablehead{\colhead{Region} & \colhead{Area} & \colhead{Net Surface Brightness} & \colhead{$\Gamma$} & \colhead{$\mathcal{N}_{-5}$} & \colhead{$\chi_\nu^2$ (d.o.f.)}  & \colhead{$L_{X,31}$} & \colhead{$F_{X,-14}$} \\ \colhead{} & \colhead{arcsec$^2$} & \colhead{cts arcsec$^{-2}$} & \colhead{} & & & \colhead{}}
\startdata
Pulsar/head & 8.96 & $70.9\pm2.8$ & $1.90\pm0.12$ & $1.1\pm0.14$ & 0.82 (54) & $8.69\pm0.50$ & $2.54_{-0.15}^{+0.12}$ \\
Lateral tails & 88.1 &  $5.84\pm0.27$ & $1.80\pm0.13$ & $0.85\pm0.12$ & 1.01 (35) & $7.50\pm0.48$ & $2.33_{-0.15}^{+0.12}$ \\
Axial tail & 23.7 & $6.45\pm0.54$ & $1.43\pm0.18$ & $0.17\pm0.04$ & 790 (2046)\tablenotemark{$\rm a$} &  $2.16\pm0.19$ & $0.81_{-0.10}^{+0.07}$ \\
Southern outflow (entire) & 36778 & $0.096 \pm 0.005$ & $1.88\pm0.06$ & $11.9\pm0.81$ & 0.90 (105) & $98.7\pm3.0$ & $29.3\pm_{-1.1}^{+1.0}$ \\
Spine & 4799 & $0.304\pm0.011$ & $2.11\pm0.07$ & $5.4\pm0.40$ & 1.01 (73) & $38.1\pm1.4$ & $9.71_{-0.43}^{+0.36}$ \\
S. outflow (excl.\ spine) & 31979 & $0.066\pm0.005$ & $1.78\pm0.09$ & $6.8\pm0.067$ & 1.09 (83) & $61.1\pm2.5$ & $19.2_{-1.1}^{+1.0}$ \\
S. outflow seg.\ 1 & 1805 & $0.434\pm0.021$  & $1.99\pm0.12$ & $1.6\pm0.22$ & 1.26 (46) & $12.7\pm0.8$ & $3.53_{-0.21}^{+0.20}$ \\
S. outflow seg.\ 2 & 2224 & $0.250\pm0.017$ & $2.28\pm0.16$ & $2.2\pm0.34$ & 1.06 (35) & $13.9\pm1.2$ & $3.14_{-0.33}^{+0.23}$ \\
S. outflow seg.\ 3 & 9463 & $0.183\pm0.008$ & $2.06\pm0.10$ & $4.6\pm0.49$ & 0.71 (42) & $33.8\pm1.8$ & $8.92_{-0.45}^{+0.59}$ \\
S. outflow seg.\ 4 & 23162 & $0.082\pm0.005$ & $1.82\pm0.12$ & $3.6\pm0.49$ & 0.95 (73) & $31.6\pm1.6$ & $9.72_{-0.65}^{+0.5}$ \\
N. outflow (entire) & 21714 & $0.081\pm0.005$ & $1.81\pm0.10$ & $4.8\pm0.55$ & 1.14 (60) & $42.0\pm2.0$ & $13.0_{-0.50}^{+0.63}$ \\
N. outflow (proximal) & 5960 & $0.080\pm0.009$ & $1.50\pm0.20$ & $0.83\pm0.20$ & 0.95 (47) & $9.74\pm0.83$ & $3.53_{-0.47}^{+0.35}$ \\
N. emission (distal) & 15754 & $0.082\pm0.006$ & $1.98\pm0.11$ & $4.3\pm0.52$ & 1.09 (67) & $33.1\pm1.8$ & $9.25_{-0.63}^{+0.56}$ \\
\vspace{-0.1cm}
\tablenotetext{}{Spectral fits for the extraction regions shown in Figures \ref{image-raw} and \ref{image-large-scale} with an absorbed PL model.  The $N_{\rm H}$  is fixed at $1.66\times 10^{22}$ cm$^{-2}$ (see text).  Unabsorbed luminosity $L_{X,31}$ and absorbed flux $F_{X,-14}$ are given for the 0.5--8 keV energy range in units of $10^{31}$ erg s$^{-1}$ and  $10^{-14}$ erg cm$^{-2}$ s$^{-1}$, respectively, for the assumed distance of 3.8 kpc. $\mathcal{N}_{-5}$ is the PL model normalization in units of $10^{-5}$ photons s$^{-1}$ cm$^{-2}$ keV$^{-1}$ at 1 keV.}
\tablenotetext{a}{C-statistic was used due to low number of counts}
\enddata
\end{deluxetable*}

\section{DISCUSSION}
In addition to the previously detected CN and southern outflow (K+08), the deep {\sl Chandra} ACIS-I observations have revealed the previously unresolved fine structure of the CN.  The overall shape of the CN and the southern outflow supports the earlier suggestion that they are due to the shocked wind of the pulsar moving in the north-northeast direction with a supersonic speed. 
The deeper image has also revealed a previously undetected faint structure, the northern outflow, extending $\approx7'$ to the north from the pulsar position.  Below we consider possible interpretations for the different components of the J1509 PWN and compare them to those seen in other PWNe.

\subsection{The Spatial and Spectral Structure of the PWN}

\subsubsection{Compact Nebula}
\begin{figure*}
\epsscale{1.1}
\plotone{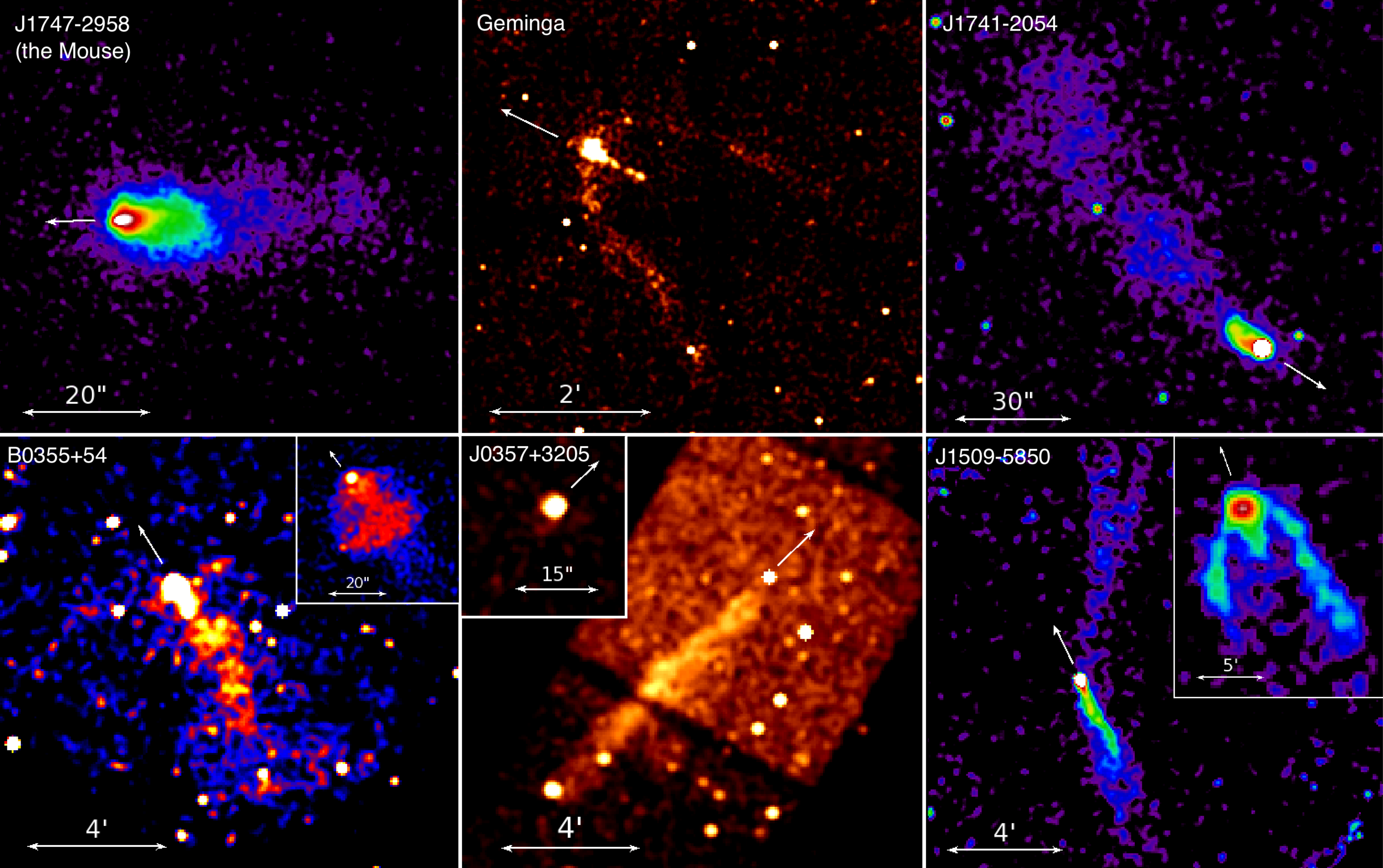}
\caption{\chan images of PWNe with prominent tails.  {\sl Top left:} J1747--2958 (the Mouse; Kargaltsev et al.\ in prep.), {\sl top center:} Geminga (Posselt et al., in prep.), {\sl top right:} J1741--2054 (Auchettl et al.\ 2015), {\sl bottom left:} B0355+54 (Klingler et al., in prep), {\sl bottom center:} J0357+3205 (De Luca et al.\ 2013), {\sl bottom right:} J1509--5850 (this paper).}
\label{bowshocks}
\end{figure*}

The structure of the J1509 CN is very similar to that of the Geminga PWN (Pavlov et al.\ 2006, 2010; see Figure \ref{bowshocks}), both in shape (one axial and two lateral tails) and in (projected) physical size ($\sim 8\times 10^{17}$ cm for the lateral tails, and $\sim 2\times 10^{17}$ cm for the axial tail, for $d_{\rm Gem} = 250$ pc and $d_{\rm J1509}=3.8$ kpc).  It is reasonable to expect that the physical sizes of compact nebulae with similar structures are proportional to $(\dot{E}/\rho v^2)^{1/2}$, where $\rho$ is the ambient density, and $v$ is the pulsar velocity. 
Assuming also similar angles $i$ between the line of sight and the pulsar motion direction, and taking into account that $\dot{E}_{\rm J1509} \simeq 15\, \dot{E}_{\rm Gem}$, we can crudely estimate the speed of the J1509 pulsar:

\begin{equation}
v_{\rm J1509}\sim790 \left( \frac{v_{\rm Gem}}{200~{\rm km~s}^{-1}} \right) \left(\frac{\rho_{\rm Gem}}{\rho_{\rm J1509}}\right)^{1/2}~{\rm km~s}^{-1}.
\label{J1509velocity}
\end{equation}
However, it is likely that Geminga lies within a local high temperature bubble, with $n<10^{-2}$ cm$^{-3}$; in contrast, J1509 is in a high density, partly neutral medium (namely the H$\alpha$ halo; see Brownsberger \& Romani 2014) with $n>1$ cm$^{-3}$.
An upper limit on the pulsar velocity follows from the lack of any significant shift of the pulsar position over $\approx$14 years, $\delta\lesssim 0\farcs5$ (see Section 3.1.1), which gives $v_{\rm J1509}\,\sin i< 640 (\delta/0.5'') (d/3.8\,{\rm kpc})$ km s$^{-1}$.  A possible lower limit on the pulsar velocity can be estimated from the lack of resolved extended emission ahead of the pulsar. 
If there is a (quasi-)isotropic component of the pulsar wind, it should produce a bullet-like termination shock with the apex (bullet head) distance $r_s = (\dot{E}/4\pi c\rho v^2)^{1/2}$.  For $\sin i=1$, the upper limit on the angular apex distance of $1\farcs0$ (see Section 3.1.1) implies $r_s <5.7\times10^{16}(d/3.8~{\rm kpc})$ cm, $p_{\rm amb} = \rho_{\rm J1509}v_{\rm J1509}^{2}>4.2\times10^{-10}(d/3.8~{\rm kpc})^{-2}$ dyn cm$^{-2}$, and $v_{\rm J1509}\gtrsim160 n^{-1/2}(d/3.8~{\rm kpc})^{-1}$ km s$^{-1}$, where $n=\rho/m_{p}$ is the ISM number density.

Despite the morphological similarities, there are also notable differences between the two PWNe. The spectra of the lateral tails are much softer in the J1509 CN than in the Geminga PWN ($\Gamma=1.8\pm0.1$ vs.\ $\Gamma=0.7$--$1.1$, respectively), while the lateral tails' luminosity, $L_{\rm 0.5-8\,keV} = (7.5\pm 0.5)\times 10^{31} (d/3.8\,{\rm kpc})^2$ erg s$^{-1}$ for J1509, is a factor of  $\sim200$ higher\footnote{The photon indices and luminosity of Geminga's lateral tails are based on recent deep {\sl Chandra} ACIS observations (Posselt et al., in prep.)}.
The higher luminosity could be attributed to the larger $\dot{E}$, but the radiative X-ray efficiency for J1509's lateral tails is also significantly higher: $\eta_{\rm 0.5-8\,keV} \equiv L_{\rm 0.5-8\,keV}/\dot{E}\simeq1.5\times10^{-4}(d/3.8\,{\rm kpc})^2$ for J1509 vs.\ $1.2\times 10^{-5} (d/0.25\,{\rm kpc})^2$ for Geminga. 
Another difference is that there is the presence of faint diffuse emission between the lateral tails (beyond the extent of the shorter axial tail) in the J1509 CN (see the inset in Figure \ref{image-raw}), while such emission is not detected in the Geminga PWN (Posselt et al., in prep.; Pavlov et al.\ 2010).

Numerical simulations of the synchrotron brightness distributions from PWNe of supersonically moving pulsars with isotropic wind ejection (Bucciantini et al.\ 2005) predict filled structures with the brightness reaching its maximum ahead of the moving pulsar and gradually decreasing behind the pulsar (see Figures 4 and 5 in Bucciantini et al.\ 2005).  
These simulated synchrotron brightness maps do not resemble the observed structures of the CNe around J1509 and Geminga.  One reason could be that the magnetic field structure is substantially different from that assumed in the simulations.  Bucciantini et al.\ (2005) mention that the magnetic field may be compressed in narrow layers near the contact discontinuity surface, which, in principle, could produce limb-brightened structures, depending on the amount of Doppler boosting (caused by a high bulk flow velocity) and the magnetic field geometry.
Another reason could be that the assumption of isotropic wind injection used in the simulations is not valid for real pulsars.  For instance, one could interpret the lateral tails as jets or an equatorial outflow bent back by the ram pressure.  However, at least the jet interpretation looks unrealistic as it requires stretching some of the parameters to their extremities. 
Indeed, a jet produced by a moving pulsar bends because the jet flow acquires a momentum from the ambient medium; it becomes comparable to the original jet flow momentum over the bending length scale (curvature radius)\footnote{This estimate neglects possible entrainment of the ambient medium into the jet.  {\bf  However, if $\xi_j \ll 1$, it is likely that the jets are inside the contact discontinuity, in which case the deflection is a result of the post-shock sweep back, and not from ballistic impact from the unperturbed ISM.} } $l_b\sim\xi_j\dot{E}/(2 c r_j \rho v^2 \sin\alpha)$, where $\xi_j<1$ is the fraction of the spin-down power $\dot{E}$ that goes into the jet, $r_j$ is the jet radius, and $\alpha$  is the angle between the initial jet direction and the pulsar's velocity.
For J1509, we obtain $l_b < 3.6 \times10^{17} \xi_j (r_j/5.7\times10^{16} {\rm cm})^{-1} (\sin \alpha)^{-1} (d/3.8\ {\rm kpc})^2 $\ cm. 
Therefore, for a jet with the angular width $\delta_j\sim1''$, a significant bending on a scale of the observed curvature radius $\delta_c\sim 1''$--$2''$ (see Section 3.1.1) implies that most of the spin-down power goes into the two jets. However, observations of PWNe with both jet and equatorial components suggest $\xi_j \ll1$ (e.g., Kargaltsev \& Pavlov 2008; Pavlov et al.\ 2003).
Therefore, the jet interpretation of the lateral tails would imply that in some pulsars jets can be more powerful than equatorial outflows.  More extensive modeling and simulations are needed to see whether it is possible to reproduce the observed CN morphologies for other assumptions on the angular distribution of the pulsar wind injection.

We also note that the morphologies of the Geminga and J1509 CNe are markedly different from those of other bow-shock PWNe.  In particular, others exhibit a CN with a local minimum in brightness in the pulsar vicinity (i.e., behind the pulsar; PWNe of PSRs J0357+3205 and J1101--6101), or exhibit a filled CN morphology (PWNe of PSRs J1747--2958, J1741--2054, and B0355+54). 
The differences, however, are not limited to X-ray morphologies only.  For instance, the radio counterparts of the PWNe associated with PSRs J1509--5850 (Ng et al.\ 2010), J1101--6101 (Pavan et al.\ 2014), and J1747--2958 (the Mouse; Yusef-Zadeh \& Gaensler  2005), the radio brightness distributions are also quite different.  While the radio brightness in the Mouse is maximal near the pulsar (similar to X-rays; Gaensler et al.\ 2004), in the other two PWNe the radio brightness is low near the pulsar and becomes higher further away (see Figure \ref{image-radio}).  
We also note that the X-ray morphologies of most of the other CNe in Figure \ref{bowshocks} do not bear much resemblance to the simulated images, except perhaps for the Mouse and J1741--2054.  Ng et al.\ (2010) mentioned an open cluster, Ruprecht 112 (C1453--623; Alter et al.\ 1970), about 4$^{\circ}$ southwest of the current pulsar position, as a possible birth place of the pulsar.  
At the distance of 3.8 kpc and the pulsar's spin-down age of 150 kyr, the required pulsar velocity would be $\sim1700$ km s$^{-1}$, much higher than the upper limit of 640 km s$^{-1}$ on the transverse velocity inferred from the lack of detectable change of the pulsar position in 14 years.
Based on the comparison of the older X-ray image with the shape of the termination shock ``bullet'' seen in numerical simulations, K+08 suggested that the pulsar velocity should be  $\sim 300$ km s$^{-1}$, which is still within the allowed velocity range (see above).  However, since the deeper X-ray images reveal substantial differences between the CN structure and the simulations, this velocity estimate should be taken with caution.

\subsubsection{Large-scale PWN Structure}
The most puzzling new feature of the J1509 PWN is the large-scale northern outflow seen up to $\sim 7'$ almost ahead of the pulsar.  One could assume that the northern outflow is a powerful relativistic jet piercing the bow shock near the pulsar.  The lack of a counter-jet could be explained by Doppler boosting, but there are other problems with the jet interpretation.
According to the estimates in Section 4.1.1, the bending length for a jet is very small, even at $\xi_j\sim 1$ and $\sin\alpha\sim 0.5$, compared to the length of the northern outflow that shows no bending.  It is also difficult to explain the relative faintness of the northern outflow closer to the pulsar 
(but outside the CN) in the jet scenario.  Finally, the X-ray radiation efficiency of the northern outflow, $\eta_{0.5-8\, keV}\sim 0.8\times 10^{-3}$, substantially exceeds the efficiencies of jets of younger PWNe with clear torus-jet morphology.  Therefore, we discard the jet interpretation for the northern outflow.

\begin{figure}
\epsscale{1.15}
\plotone{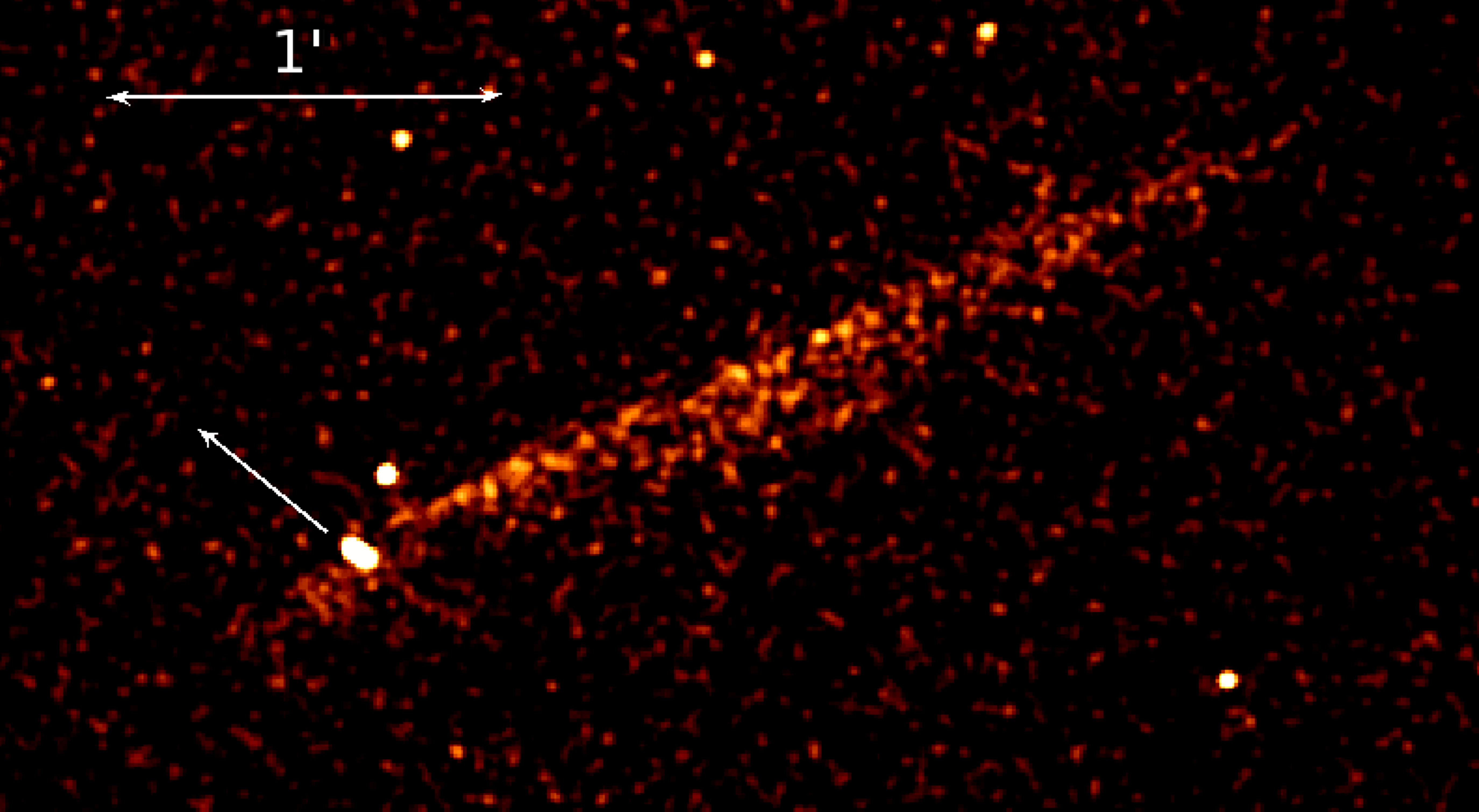}
\caption{Merged image based on 194 ks of \chan ACIS-S observations of PSR B2224+65 (the Guitar Nebula). The white arrow shows the direction of the pulsar's proper motion.}
\label{image-guitar}
\end{figure}

Similar arguments have led Bandiera (2008) to conclude that the extended X-ray feature accompanying the Guitar nebula (which also exhibits a large-scale outflow misaligned with the pulsar motion direction; see Figure \ref{image-guitar} and Johnson \& Wang 2010) cannot be interpreted as a hydrodynamical jet.  Following Bandiera (2008), we consider a scenario in which particles with the highest energies leak from the termination shock apex into the ISM, and travel along the ordered ISM magnetic field.
Bandiera (2008) suggested that the length of the Guitar PWN feature, $\sim 1.4$ pc, is on the order of the radiation length (i.e., the distance over which particles lose a substantial fraction of their energy to synchrotron radiation).  However, this assumption implies a propagation speed of only a few thousands km s$^{-1}$, instead of the speed of light, as expected for particles with large (bulk) Lorentz factors.     
Bandiera (2008) argued that the turbulent component of the magnetic field would provide a scattering mechanism and slow down particle propagation (which can then be considered as diffusion) along the ISM magnetic field to the required non-relativistic speed.  The observed hint of softening of the X-ray spectrum in J1509's northern outflow can be considered as supporting this scenario.  
A problem with this scenario is that in all three cases with prominent misaligned (with respect to pulsar motion) X-ray features (the Guitar, J1509, and J1101--6101), the features appear to be one-sided or show very faint ``counter-features'', which means that there is a preferred direction for the escape of the high-energy particles. The cause of this asymmetry is currently unclear. 
On the other hand, since the particles escaping from the termination shock apex into the ISM must be the most energetic ones (those with large gyration radii; see Bandiera 2008), the lack of radio synchrotron emission can be naturally explained by the deficit of lower energy particles.

The southern outflow is aligned with the CN symmetry axis and clearly represents a continuation of the CN (see Figure \ref{image-mid-scale}).  Therefore, we can assume that at least the brighter part of the southern outflow can be interpreted as a tail filled with the shocked pulsar wind confined by the ram pressure.
The lack of evidence for synchrotron burn-off in the X-ray spectra suggests a high flow speed and/or a low magnetic field.  Alternatively, there could be some ``reheating'' due to in-situ conversion of magnetic field energy into particle energy, e.g., via turbulent processes and accompanying reconnection, which might explain the hint of spectral hardening at large distances from the pulsar (see Figure \ref{image-gamma-fn-dist}.)  
Indeed, Ng et al.\ (2010) reported the presence of substantial turbulent magnetic field components in the tail using radio polarimetry data.  

There are, however, some puzzling aspects associated with the southern outflow.  One is that the radio emission brightens with distance from the pulsar and becomes broader than the X-ray emission beyond $\approx 3'$ (see Figure \ref{image-radio}).  Such behavior could possibly be caused by entrainment of ambient matter (see also Morlino et al.~2015), which decreases the flow speed, increases the transverse size of the tail, and leads to adiabatic cooling of the outflowing matter. 
This tentative explanation remains to be checked with detailed modeling of entrainment effects.

Another puzzling aspect is the similarity of the wedge-like shapes of the fainter part of the southern outflow (i.e., the southern outflow excluding the spine) and the northern outflow.  Given the qualitative differences in the above interpretations of the northern and southern outflows, these similarities could be attributed to merely a chance coincidence.
Alternatively, the structural and spectral similarities could be explained if both the northern emission and the southern emission (excluding the spine) are caused by very energetic particles that leaked from the termination shock apex in opposite directions and diffused along the curved ISM magnetic field (the curvature of the magnetic field is required to explain the different orientations of the ``wedges''). 
In this scenario, the spine (whose spectrum is slightly softer than that of the rest of the southern outflow) is still interpreted as the pulsar tail, coincidentally projected onto the sky within the southern outflow.  This interpretation may look somewhat artificial, but it cannot be firmly excluded with the currently available data.

The spectra of the large-scale outflows measured in several regions allow one to crudely estimate magnetic fields in these regions.  For a PL synchrotron spectrum with photon index $\Gamma$, the magnetic field at a given magnetization parameter $\sigma= w_B/w_e$ (ratio of energy densities of magnetic field and relativistic electrons) depends on the ratio of the luminosity $L(\nu_m,\nu_M)$, measured in the  $\nu_m <\nu <\nu_M$ frequency range, to the radiating volume $V$. 
A simple generalization of Equation (7.14) in Pacholczyk (1970)\footnote{Equation (7.14) gives the magnetic field for $\sigma=3/4$ (which minimizes the total energy density $w_B+w_e$) and assumes $\nu_m=\nu_1$ and $\nu_M=\nu_2$.} to the case of arbitrary $\sigma$ gives
\begin{equation}
B=\left[\frac{L(\nu_m,\nu_M)\sigma}{\mathcal{A} V} \frac{\Gamma-2}{\Gamma-1.5}\frac{\nu_1^{1.5-\Gamma}-\nu_2^{1.5-\Gamma}}{\nu_m^{2-\Gamma}-\nu_M^{2-\Gamma}} \right]^{2/7}.
\label{synch_magn_field}
\end{equation}
In this equation, $\nu_1$ and $\nu_2$  are the characteristic synchrotron frequencies ($\nu_{\rm syn}\simeq eB\gamma^2/2\pi mc$) corresponding to the boundary energies ($\gamma_1 m_ec^2$ and $\gamma_2 m_ec^2$) of the electron spectrum ($dN_e/d\gamma \propto \gamma^{-p}\propto \gamma^{-2\Gamma+1}$; $\gamma_1<\gamma<\gamma_2$), 
and $\mathcal{A}=2^{1/2}e^{7/2}/18 \pi^{1/2}m_e^{5/2}c^{9/2}$.  The photon indices and 0.5--8 keV luminosities have been measured from the spectral fits (see Table \ref{spatially-resolved-spectra}).
We estimate the radiating volumes from the \chan images assuming conical shapes of the outflows. Since $L\propto d^2$ and $V\propto d^3$, the magnetic field estimated from Equation (\ref{synch_magn_field}) weakly depends on the assumed distance, $B\propto d^{-2/7}$.  The boundary frequencies $\nu_1$ and $\nu_2$ are unknown. 
The exact value of $\nu_2$ is not really important as long as $(\nu_1/\nu_2)^{\Gamma -1.5}\ll 1$, so we choose a plausible (but arbitrary) value $h\nu_2=100$ keV.  The unknown frequency $\nu_1$ is the main source of uncertainty of the magnetic field at a given $\sigma$, for the measured spectral slopes. 
For the southern outflow, we can at least estimate a lower limit on $\nu_1$ from the requirement that extrapolation of the X-ray PL spectrum to lower frequencies does not exceed the harder radio spectrum ($\Gamma_{\rm rad}=1.26\pm 0.04$; Ng et al.\ 2010), but even the limit, $\nu_1\gtrsim 10^{10}$--$10^{12}$ Hz, is not very certain because of the uncertainty of the X-ray photon indices (and because the slope of the radio spectrum was measured in a broader region).
Since the northern outflow was not detected in the radio, we can only guess that $\nu_1$ is higher there than in the southern outflow (which would be consistent with our hypothesis that the northern outflow emission is due to leaked particles with very high energies). Given these uncertainties, we estimate the magnetic fields assuming $\nu_1=100$ GHz for all the regions. If future multiwavelength observations yield different estimates of $\nu_1$, the magnetic field could be scaled (approximately) as $B\propto \nu_1^{(3-2\Gamma)/7}$ (if $(\nu_1/\nu_2)^{\Gamma -1.5}\ll 1$). 

The equipartition magnetic fields, $B_{\rm eq}=B\sigma^{-2/7}$, for different parts of the J1509 southern and northern outflows are given in Table \ref{b-fields}, for the above-described choice of boundary frequencies and $d=3.8$ kpc.  Although the actual magnetic field strengths can differ from those in Table \ref{b-fields} by a factor of a few, we can at least hope that their relative values reflect the spatial dependence of the magnetic field in the emitting regions.
Surprisingly, the magnetic field in the southern outflow shows a hint of non-monotonic dependence on distance from the pulsar: the field grows from segment 1 to segment 2, and then decreases outwards (in concert with the photon index $\Gamma$ and slope $p$ of electron spectrum).  The reason for such behavior is unclear, and we cannot exclude that it is simply due to a variation of unknown parameters (e.g., $\nu_1$ and/or $\sigma$ along the outflow). 

Since the southern outflow was detected in radio (Ng et al.\ 2010), the radio spectrum ($\Gamma_{\rm rad}\approx 1.26$) can also be used for magnetic field estimates. If the radio and X-ray spectra are generated in different regions (e.g., the radio emission comes from a  ``sheath'' around an X-ray emitting ``spine'' of the outflow), the magnetic field in the radio-emitting region should be estimated solely from the radio spectrum, which gives $B_{\rm rad} \simeq 13\, \sigma_{\rm rad}^{2/7}$ $\mu$G for $L_{\rm rad}(\nu_m,\nu_M) = 1.5\times 10^{32}$ erg s$^{-1}$, $V_{\rm rad} = 5\times 10^{56}$ cm$^3$, $\nu_1=\nu_m=10$ MHz, $\nu_2=\nu_M=100$ GHz ($\sigma_{\rm rad}$ is the magnetization in this region).
This value of $B_{\rm rad}$ is close to those determined from the X-ray spectra in segments 3 and 4 (assuming $\sigma_{\rm rad} \sim \sigma$).  If the radio spectrum in the X-ray detected part of the southern outflow is generated in the same region as the X-ray emission, then the mutiwavelength (radio to X-rays) spectrum can be considered as a broken PL. 
The magnetic field from such a spectrum could be estimated for a given magnetization $\sigma = (\sigma_{\rm rad}^{-1} + \sigma_X^{-1})^{-1}$, where $\sigma_{\rm rad}$ and $\sigma_X$ (each of them is greater than $\sigma$) are magnetizations corresponding to the energy densities of particles responsible for radiation below and above the break frequency, respectively.  In this case, the equipartition field would be slightly higher than those estimated separately from the X-ray and radio spectra, but the difference would be within the uncertainties.

According to Table \ref{b-fields}, the difference of the equipartition fields in the two segments of the northern outflow is within the $B_{\rm eq}$ uncertainties caused by the uncertainties of the spectral slopes. The characteristic equipartition field, $B_{\rm eq}\sim 5$--10 $\mu$G, is lower than typical values for the southern outflow\footnote{It would be even lower if we assume a $\nu_1$ value higher than 100 GHz for the northern outflow.}. 
However, if the X-ray emission from the northern outflow is indeed due to particles of very high energy leaked from the termination shock apex, then  
$\sigma$ can be very different from unity, and $B$ very different from $B_{\rm eq}$.

The magnetic field strengths can be used to estimate the Lorentz factors,  gyration radii, and characteristic synchrotron cooling times of electrons radiating synchrotron photons with energy $E$:

\begin{equation}
\gamma\sim 10^{8}(E/1\,{\rm keV})^{1/2}(B/20\,\mu{\rm G})^{-1/2} ,
\end{equation}
 
\begin{equation}
r_g\sim  10^{14}(E/1\,{\rm keV})^{1/2}(B/20\,\mu{\rm G})^{-3/2}~{\rm cm},
\end{equation}

and 
\begin{equation}
t_{\rm syn}\sim 400 (E/1\,{\rm keV})^{-1/2}(B/20\,\mu{\rm G})^{-3/2}~{\rm yr}\,.
\end{equation}

The spatially-resolved X-ray spectra of the outflows allow one to crudely estimate (or constrain) characteristic flow velocities.  For instance, if the apparent spectral softening in the northern outflow is real, the flow velocity can be estimated as $v_{\rm flow} \sim l/\tau_{\rm syn}\sim 3000 (l/6\,{\rm pc}) (B/7\,\mu{\rm G})^{3/2}$ km s$^{-1}$, where $l$ is the spatial scale of the alleged softening. 
This is, of course, just an order-of-magnitude estimate, but the fact that $v_{\rm flow} \ll c$ is consistent with the scenario by Bandiera (2008), which suggests that propagation of the leaked high-energy particles is slowed down by diffusion in a turbulent magnetic field (see above). 

Knowing the equipartition magnetic field and the flow velocity, we can estimate the energy density,
\begin{eqnarray}
w & = & w_B + w_e = [B_{\rm eq}^2/(8\pi)](\sigma^{4/7}+\sigma^{-3/7})^2 \\
\nonumber
 & \sim & 2\times 10^{-12} (B_{\rm eq}/7\,\mu{\rm G})^2 (\sigma^{4/7}+\sigma^{-3/7})^2\,\,{\rm erg}\,{\rm cm}^{-3}, 
\label{energy_density_north}
\end{eqnarray}
and the energy transfer rate ($\sim$ energy injection rate),
\begin{eqnarray}
\dot{E}_{\rm North} & \sim & (w+p) S v_{\rm flow} 
\label{energy_injection_north}\\
\nonumber
& \sim & 4\times 10^{33} (B_{\rm eq}/7\,\mu{\rm G})^2 (S/3\times 10^{36}\, {\rm cm}^2)\\
\nonumber
&\times & (v_{\rm flow}/3000\,{\rm km}\,{\rm s}^{-1}) (\sigma^{4/7}+(2/3)\sigma^{-3/7})^2\,\,{\rm erg}\,{\rm s}^{-1} ,
\end{eqnarray}
for the northern outflow, {\bf where $w+p$ is the enthalpy density, $p=w_B+w_e/3$ is the pressure of the relativistic particles and magnetic field, and $S$ is the characteristic cross-sectional area.} 
As expected, $\dot{E}_{\rm North}$ is a small fraction of the pulsar's spin-down power ($\dot{E}_{\rm North}\sim 10^{-2}\dot{E}$ for $\sigma\sim 1$).  
On the other hand, $\dot{E}_{\rm North}$ exceeds the 0.5--8 keV luminosity of the northern outflow (by a factor of 7 at the parameters chosen and $\sigma\sim 1$, but the value of this factor strongly depends on the poorly known outflow parameters).

In the southern outflow we see no spectral softening with increasing distance from the pulsar (i.e., no synchrotron cooling).  If there is no in-situ particle acceleration in the outflow, the lack of cooling sets a lower limit on the flow velocity of a few thousand km s$^{-1}$ (K+08), well above the upper limit on the pulsar velocity.  The value of the limiting speed is somewhat uncertain because of the magnetic field uncertainties.  
Moreover, the hint of spectral hardening with increasing distance from the pulsar (see Figure \ref{image-gamma-fn-dist}) suggests that particle acceleration may occur in the southern outflow (e.g., due to magnetic field reconnection), which would invalidate the velocity estimates based on passive cooling.  
An upper limit of a few tens of thousands km s$^{-1}$ on the characteristic flow velocity can be estimated from the condition $\dot{E}_{\rm South}< \dot{E}$, where $\dot{E}_{\rm South}\sim (w+p) S v_{\rm flow}$, similar to Equation (7).

\begin{deluxetable}{lcccc}[H]
\tablecolumns{5}
\tablecaption{Equipartition Magnetic Field Estimates
\label{b-fields}}
\tablewidth{0pt}
\tablehead{\colhead{Region}  & \colhead{$p$} &  \colhead{$L_{32}$} & \colhead{$V_{55}$}  & \colhead{$B_{\rm eq,-6}$} }
\startdata
S.\ outflow seg.\ 1 & $2.98\pm0.24$ &1.3 & 2.6 & $36_{-12}^{+23}$ \\
S.\ outflow seg.\ 2 & $3.56\pm 0.32$ &  1.4 & 7.4 & $89_{-42}^{+16}$  \\ 
S.\ outflow seg.\ 3 & $3.12\pm0.20$ & 3.4 & 42 & $29_{-9}^{+14}$ \\ 
S.\ outflow seg.\ 4  & $2.64\pm0.24$   & 3.2 & 125 & $9_{-3}^{+4}$ \\
N.\ outflow (proximal) & $2.00\pm0.40$ &  1.0 & 9.1 & $5_{-1}^{+4}$ \\ 
N.\ outflow (distal) & $2.96\pm0.22$ &  3.3 & 76 & $10_{-2}^{+4}$ \\
\vspace{-0.3cm}
\tablecomments{The values are given for different regions of the northern (N) and southern (S) outflows shown in Figure~\ref{image-large-scale}.  The parameter $p$ is the slope of electron spectrum, $V_{55}$ is the volume of the corresponding segment  in units of $10^{55}$ cm$^{3}$, $L_{32}$ is the 0.5--8 keV luminosity in units of $10^{32}$ erg s$^{-1}$ (for $d=3.8$ kpc), $B_{\rm eq, -6}$ is the equipartition magnetic field in units of $\mu$G. The uncertainties of $B_{\rm eq,-6}$ in the last column are due to the X-ray PL slope uncertainties, they do not include the uncertainty of the boundary frequencies fixed at $\nu_1=10^{11}$ Hz and $h\nu_2 = 100$ keV (see text). }
\enddata
\end{deluxetable}

\begin{figure}
\epsscale{1.15}
\plotone{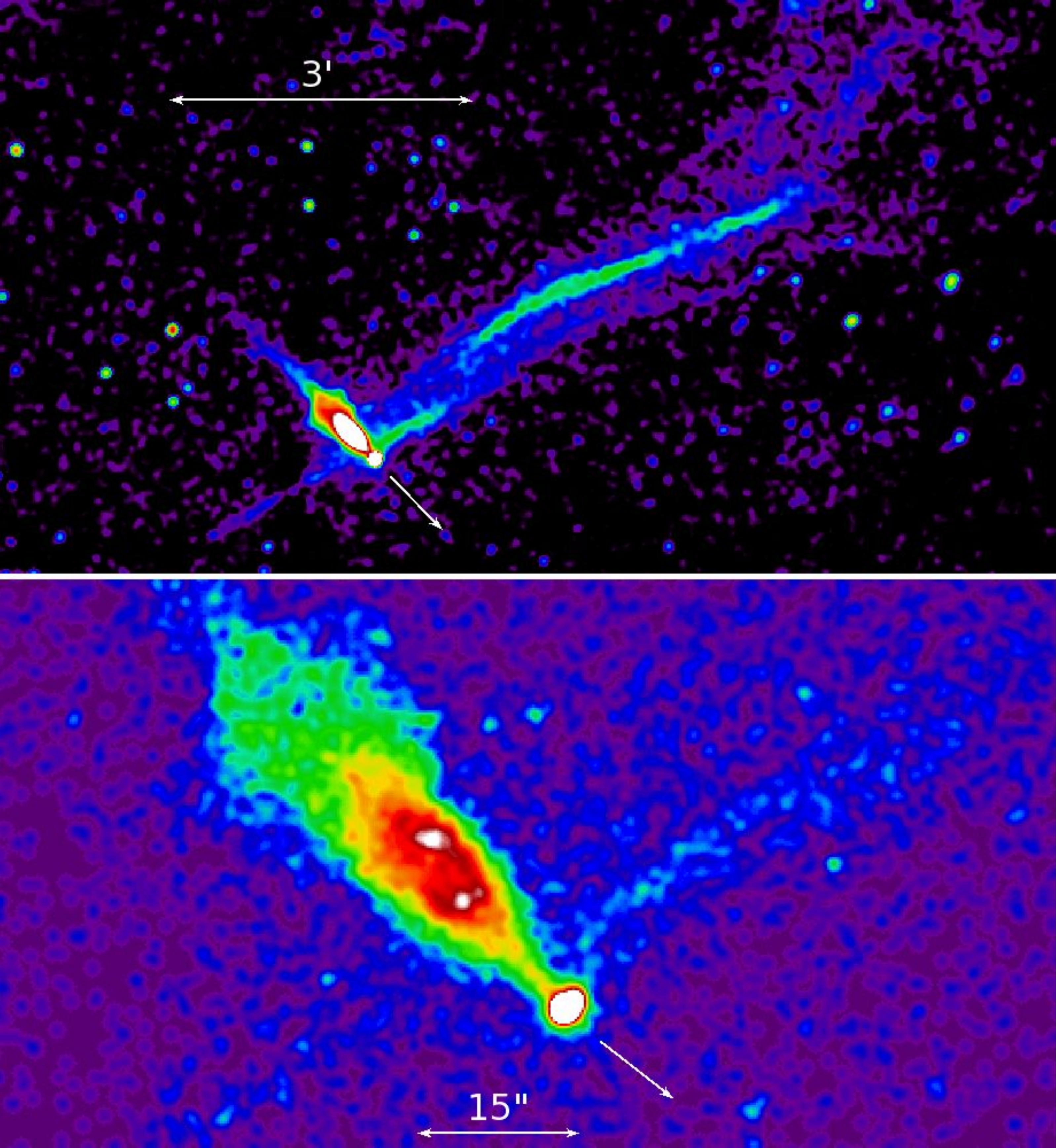}
\caption{Images of PSR J1101--6101 and its PWN obtained from merged \chan ACIS-I observations (300 ks total exposure, 0.5--8.0 keV band; see Pavan et al.\ 2015).  The top panel shows an image of the large-scale structure, binned by a factor of 2 (1$''$ pixel size).  The bottom panel shows an image of the small-scale structure (the image pixel size is 0.25$''$).  Both images are smoothed with a 3-pixel Gaussian kernel.  The white arrows represent the assumed direction of the pulsar's proper motion (Tomsick et al.\ 2012).}
\label{image-lighthouse}
\end{figure}

As we mentioned above, the J1509 and Guitar PWNe are not the only ones which exhibit puzzling asymmetric structures misaligned with their apparent pulsar motion  directions.  Two other examples are the outflows associated with PSR J1101--6101 (the Lighthouse Nebula; Pavan et al.\ 2014) and PSR J2055+2539 (Marelli et al.\ 2015).  The former one shows a particularly remarkable structure.
The deep 300 ks \chan ACIS image of the Lighthouse Nebula shown in Figure \ref{image-lighthouse} reveals a long ($\sim 11$ pc) jet-like outflow that originates near the pulsar, bends back in the direction opposite to that of the pulsars' proper motion, and then sharply turns in a direction nearly orthogonal to the proper motion direction (see also Pavan et al.\ 2015).  
This outflow can possibly be interpreted in the framework of the Bandiera (2008) scenario, similar to Guitar and J1509 misaligned outflows.  
{\bf In this scenario, the external (ISM) magnetic field lines are illuminated by the escaping wind particles traveling along the distorted magnetic field (cf.\ magnetic ``draping'';  Lyutikov 2006 and Dursi \& Pfrommer 2008).}
Additionally, the PWN of PSR B0355+54 also shows hints of a misaligned outflow (see Figure \ref{bowshocks}), although in this case it is rather faint and is seen on both sides of the bow shock (perhaps due to the lower speed of the pulsar).  An in-depth discussion of this PWN will be given in a separate paper (Klingler et al., in prep.).

\section{Summary}
The most important findings from our deep \chan ACIS-I observations of J1509 can be summarized as follows.

We were able to resolve the fine structure of the CN in the immediate pulsar vicinity.  
The CN consists of two lateral tails, apparently emerging from the pulsar and resolved up to 0.2 pc behind the pulsar, and a short, 0.05 pc, axial tail along the CN symmetry axis. 
The CN structure and size are remarkably similar to those of the Geminga, PWN but different from several other bow shock PWNe. 
Such PWN morphology challenges the existing numerical simulations of synchrotron emission from the PWNe of supersonically moving pulsars.
The morphological differences from other PWNe could be attributed to different initial wind magnetizations or, more likely, to anisotropic wind ejection, which affects the geometry of the magnetic field downstream of the termination shock and the PWN morphology. 

We found that the CN narrows beyond 0.2--0.3 pc, reaching the minimum width (where the lateral tails apparently merge) at about 0.5 pc behind the pulsar. 
Beyond this distance the flow expands and forms the diverging 7 pc long southern outflow, discovered by K+08 who interpreted it as a tail of shocked pulsar wind confined by the ram pressure of the ambient medium. 
The PL spectra of the southern outflow do not show softening (synchrotron cooling) with increasing distance from the pulsar, which is expected for such long tails.
The lack of softening (with a hint of hardening) may suggest that particle acceleration occurs within the tail. 

We also discovered a previously undetected wedge-like northern outflow, seen up to about 6 pc, with the wedge axis inclined at an angle of about $33^\circ$ from the pulsar's proper motion direction, inferred from the CN symmetry axis.
This outflow is possibly formed by energetic pulsar wind particles escaping from the apex of the bow shock and diffusing along the ISM magnetic field. 
The equipartion magnetic field in this outflow is about a few $\mu$G, the flow speed is a few thousand km s$^{-1}$, and the energy injection rate is $\gtrsim 10^{33}$ erg s$^{-1}$.
The discovery of the northern outflow puts J1509 into the (so far) small group of supersonically moving pulsars with misaligned outflows (i.e., outflows inclined to the pulsar trajectory at substantial angles). 
To better understand the nature of the misaligned outflows, further deep multiwavelength observations with high angular resolution are needed.

{\em Facilities:} \facility{{\sl CXO, XMM-Newton}}

\acknowledgements

We would like to thank Maxim Lyutikov, Andrei Bykov, Giovanni Morlino, Marina Romanova, and John Hewitt for the very insightful discussions, as well as our referee, Roger Romani, for the helpful suggestions.  Support for this work was provided by the National Aeronautics and Space Administration through {\sl Chandra} Award Number G03-14082  issued by the {\sl Chandra} X-ray Observatory Center, which is operated by the Smithsonian Astrophysical Observatory for and on behalf of the National Aeronautics Space Administration under contract NAS8-03060. The work was also partly supported by NASA grant NNX08AD71G.


\begin{thebibliography}{ }

\bibitem[Abdo et al.(2013)]{2013ApJS..208...17A} Abdo, A.~A., Ajello, M., Allafort, A., et al.\ 2013, \apjs, 208, 17 

\bibitem[Auchettl et al.(2015)]{2015ApJ...802...68A} Auchettl, K., Slane, P., Romani, R.~W., et al.\ 2015, \apj, 802, 68 

\bibitem[Bandiera(2008)]{2008A&A...490L...3B} Bandiera, R.\ 2008, \aap, 490, L3 

\bibitem[Brownsberger \& Romani(2014)]{2014ApJ...784..154B} Brownsberger, S., \& Romani, R.~W.\ 2014, \apj, 784, 154 

\bibitem[Bucciantini et al.(2005)]{2005A&A...434..189B} Bucciantini, N., Amato, E., \& Del Zanna, L.\ 2005, \aap, 434, 189 

\bibitem[De Luca et al.(2011)]{2011ApJ...733..104D} De Luca, A., Marelli, M., Mignani, R.~P., et al.\ 2011, \apj, 733, 104 

\bibitem[Dursi \& Pfrommer(2008)]{2008ApJ...677..993D} Dursi, L.~J., \& Pfrommer, C.\ 2008, \apj, 677, 993 

\bibitem[Freeman et al.(2002)]{2002ApJS..138..185F} Freeman, P.~E., Kashyap, V., Rosner, R., \& Lamb, D.~Q.\ 2002, \apjs, 138, 185 

\bibitem[Gaensler et al.(2004)]{2004ApJ...616..383G} Gaensler, B.~M., van der Swaluw, E., Camilo, F., et al.\ 2004, \apj, 616, 383 

\bibitem[Hui \& Becker(2007)]{2007A&A...467.1209H} Hui, C.~Y., \& Becker, W.\ 2007, \aap, 467, 1209 

\bibitem[Johnson \& Wang(2010)]{2010MNRAS.408.1216J} Johnson, S.~P., \& Wang, Q.~D.\ 2010, \mnras, 408, 1216 

\bibitem[Kargaltsev \& Pavlov(2008)]{2008AIPC..983..171K} Kargaltsev, O., \& Pavlov, G.~G.\ 2008, 40 Years of Pulsars: Millisecond Pulsars, Magnetars and More, 983, 171 

\bibitem[Kargaltsev et al.(2008)]{2008ApJ...684..542K} Kargaltsev, O., Misanovic, Z., Pavlov, G.~G., Wong, J.~A., \& Garmire, G.~P.\ 2008, \apj, 684, 542

\bibitem[Kramer et al.(2003)]{2003MNRAS.342.1299K} Kramer, M., Bell, J.~F., Manchester, R.~N., et al.\ 2003, \mnras, 342, 1299 

\bibitem[Lyutikov(2006)]{2006MNRAS.373...73L} Lyutikov, M.\ 2006, \mnras, 373, 73 

\bibitem[Marelli et al.(2015)]{2015arXiv150905833M} Marelli, M., Pizzocaro, D., De Luca, A., et al.\ 2015, arXiv:1509.05833 

\bibitem[McGowan et al.(2006)]{2006ApJ...647.1300M} McGowan, K.~E., Vestrand, W.~T., Kennea, J.~A., et al.\ 2006, \apj, 647, 1300 

\bibitem[Morlino et al.(2015)]{2015MNRAS.454.3886M} Morlino, G., Lyutikov, M., \& Vorster, M.\ 2015, \mnras, 454, 3886 

\bibitem[Morrison \& McCammon(1983)]{1983ApJ...270..119M} Morrison, R., \& McCammon, D.\ 1983, \apj, 270, 119

\bibitem[Ng et al.(2010)]{2010ApJ...712..596N} Ng, C.-Y., Gaensler, B.~M., Chatterjee, S., \& Johnston, S.\ 2010, \apj, 712, 596 

\bibitem[Pacholczyk(1970)]{1970ranp.book.....P} Pacholczyk, A.~G.\ 1970, Series of Books in Astronomy and Astrophysics, San Francisco: Freeman, 1970,  

\bibitem[Pavan et al.(2014)]{2014A&A...562A.122P} Pavan, L., Bordas, P., P{\"u}hlhofer, G., et al.\ 2014, \aap, 562, A122

\bibitem[Pavan et al.(2015)]{2015arXiv151101944P} Pavan, L., P{\"u}hlhofer, G., Bordas, P., et al.\ 2015, arXiv:1511.01944 

\bibitem[Pavlov et al.(2003)]{2003ApJ...591.1157P} Pavlov, G.~G., Teter, M.~A., Kargaltsev, O., \& Sanwal, D.\ 2003, \apj, 591, 1157 

\bibitem[Pavlov et al.(2006)]{2006ApJ...643.1146P} Pavlov, G.~G., Sanwal, D., \& Zavlin, V.~E.\ 2006, \apj, 643, 1146 

\bibitem[Pavlov et al.(2010)]{2010ApJ...715...66P} Pavlov, G.~G., Bhattacharyya, S., \& Zavlin, V.~E.\ 2010, \apj, 715, 66 

\bibitem[Taylor \& Cordes(1993)]{1993ApJ...411..674T} Taylor, J.~H., \& Cordes, J.~M.\ 1993, \apj, 411, 674 

\bibitem[Tomsick et al.(2012)]{2012ApJ...750L..39T} Tomsick, J.~A., Bodaghee, A., Rodriguez, J., et al.\ 2012, \apjl, 750, L39 

\bibitem[Weltevrede et al.(2010)]{2010ApJ...708.1426W} Weltevrede, P., Abdo, A.~A., Ackermann, M., et al.\ 2010, \apj, 708, 1426 

\bibitem[Yusef-Zadeh \& Gaensler(2005)]{2005AdSpR..35.1129Y} Yusef-Zadeh, F., \& Gaensler, B.~M.\ 2005, Advances in Space Research, 35, 1129 

\end{thebibliography}
\end{document}